\documentclass[a4paper,11pt]{article}
\pdfoutput=1

\usepackage{jheppub}
\usepackage[T1]{fontenc} \usepackage{dsfont}

\usepackage{comment}
\usepackage{stackengine,xcolor}
\usepackage{Settings}
\usepackage{blkarray}\usepackage{todonotes}
\usepackage{subcaption}
\usepackage{tabularx}
\usepackage{mathbbol}

\DeclareMathOperator\arccoth{arccoth}

\usepackage{array, cellspace, makecell, caption, booktabs}
\setcellgapes{5pt}
\newcommand\mathcalbb[2][2]{\stackengine{0pt}{$\color{black!30}\mathcal{#2}$}{$\mkern#1mu\mathcal{#2}$}{O}{l}{F}{F}{L}}

\newcommand{\badat}{\begin{alignedat}}
 \newcommand{\eadat}{\end{alignedat}}

\def\lambdabar{\protect\@lambdabar}
\def\@lambdabar{\relax \bgroup
\def\@tempa{\hbox{\raise.73\ht0
\hbox to0pt{\kern.2\wd0\vrule width.7\wd0
height.1pt depth.1pt\hss}\box0}}\mathchoice{\setbox0\hbox{$\displaystyle\lambda$}\@tempa}{\setbox0\hbox{$\textstyle\lambda$}\@tempa}{\setbox0\hbox{$\scriptstyle\lambda$}\@tempa}{\setbox0\hbox{$\scriptscriptstyle\lambda$}\@tempa}\egroup}

\newcommand{\BETA}{\gamma}

\usepackage[colorlinks = true,
            linkcolor = blue,
            urlcolor  = blue,
            citecolor = blue,
            anchorcolor = blue]{hyperref}

\title{Wave scattering event shapes at high energies}
\date{\today}

\author[1]{Riccardo Gonzo}
\author[1]{and Anton Ilderton}

\affiliation[1]{Higgs Centre for Theoretical Physics, School of Physics and Astronomy, The University of Edinburgh, EH9 3FD, Scotland}

\emailAdd{rgonzo@ed.ac.uk}\emailAdd{anton.ilderton@ed.ac.uk}

\abstract{We study the space and properties of global and local observables for radiation emitted in the scattering of a massive scalar field in gauge and gravitational plane-wave backgrounds, in both the quantum and classical theory. We first compute the radiated momentum and angular momentum flow, demonstrating that they are good local observables determined by the amplitude and phase of the waveform.
We then focus on the corresponding global observables, which in the gravitational case requires dealing with the collinear divergence of the gravitational Compton cross-section. We show using the KLN theorem that we can obtain an infrared-finite cross-section only by summing over forward scattering diagrams; this suggests dressing the initial state in the direction collinear to the plane wave in order to be able to compute observables integrated over the celestial sphere. Finally, we explore the high-energy behaviour of our observables. We find that classical global observables generically exhibit a power-law mass divergence in electrodynamics and a logarithmic mass divergence in gravity, even when radiation reaction is included. We then show explicitly how this is consistently resolved in the full quantum theory.}

\newcommand{\ud}{\mathrm{d}}
\newcommand{\LCm}{{\scriptscriptstyle -}}
\newcommand{\LCp}{{\scriptscriptstyle +}}

\newcommand{\LCperp}{{\scriptscriptstyle \perp}}
\newcommand{\be}{\begin{equation}}
\newcommand{\ee}{\end{equation}}
\newcommand{\sfx}{{\mathsf x}}

\begin{document}

\clearpage
\maketitle
\section{Introduction and motivation}
\label{sec:intro}

Recent years have seen a resurgence of activity in computing classical observables from scattering amplitudes, motivated by interest in gravitational wave physics. There are many open questions to address in understanding the \textit{space} and \textit{properties} of on-shell scattering observables in both the classical and quantum regimes. One can ask, for example, what are the roles of local vs.~global observables and their relation with scattering dynamics? Do these observables have a smooth high-energy limit? 

We consider here the space of on-shell radiative observables in wave-particle scattering. Global observables like the cross-section, impulse and angular impulse have been extensively studied but, despite this, a gauge-invariant definition of the angular momentum operator was lacking until recently~\cite{Manohar:2022dea,DiVecchia:2022owy}; it would be interesting to better understand its properties~\cite{Riva:2023xxm}. Furthermore, local observables\footnote{In this paper, \emph{local} is synonymous with \emph{differential} and refers to the dependence of observables on the angles of the celestial sphere.} like the differential cross-section and the waveform contain, by definition, more information than their global counterparts. This begs the question of whether the differential linear and angular momentum, sometimes referred to as  `flows' \cite{Basham:1977iq,Basham:1978bw,Basham:1978zq,Korchemsky:1999kt,Belitsky:2001ij,Berger:2003iw,Belitsky:2013bja,Belitsky:2013xxa}, are proper local observables. For the former case, it was shown in~\cite{Cristofoli:2021vyo} that the linear momentum flow is directly related to the amplitude of the waveform. In this paper we will show that angular momentum flow is related to the \emph{phase} of the waveform. 

Other observables for the wave scattering problem have been studied with modern on-shell techniques, mostly working in the geometric optics approximation in order to make contact with the eikonal expansion~\cite{Bjerrum-Bohr:2014zsa,Bjerrum-Bohr:2016hpa,Bastianelli:2021nbs,AccettulliHuber:2020oou,Bellazzini:2022wzv} but also beyond~\cite{Cristofoli:2021vyo,Chen:2022yxw,Bautista:2021wfy}. In particular, quantum long-range effects have been taken into account within an effective field theory approach to gravity, with interesting consequences for the equivalence principle~\cite{Bjerrum-Bohr:2016hpa,Bai:2016ivl,Chi:2019owc,Brandhuber:2019qpg,Kim:2022iub}. 

Turning to the study of classical wave observables, one can ask i) what is their regime of validity in the full theory, ii) how to perform a consistent resummation of results obtained in perturbation theory and iii) how to resolve divergences when they are encountered. Recent work on the eikonal operator has largely focused on the first two problems ~\cite{DiVecchia:2022nna,Cristofoli:2021jas,DiVecchia:2022piu} in the context of the classical two-body problem for a pair of massive point particles emitting radiation. These questions have in particular been investigated also in the ultrarelativistic limit where we can map the problem to the motion of one particle in the point-like background generated by the other (i.e.~a shockwave)~\cite{DEath:1976bbo,Gruzinov:2014moa,Adamo:2021rfq}; see also~\cite{Ciafaloni:2015vsa,Ciafaloni:2015xsr,Ciafaloni:2016nul,Ciafaloni:2018uwe}. Regarding the third problem, recent work on the classical two-body problem both in electrodynamics \cite{Saketh:2021sri,Bern:2021xze} and in general relativity \cite{Herrmann:2021tct,Mougiakakos:2021ckm,Jakobsen:2021smu,DiVecchia:2021bdo,DiVecchia:2022nna,Dlapa:2022lmu,Dlapa:2023hsl,Bini:2022enm,Damour:2022ybd} has shown that classical radiative observables may fail to have a smooth high-energy limit: the radiated momentum, for example, exhibits a power-law mass divergence in electrodynamics and a logarithmic mass divergence in general relativity. In this context, it has been discussed in~\cite{Kovacs:1977uw,Kovacs:1978eu,DEath:1976bbo,Gruzinov:2014moa,DiVecchia:2022nna} how non-perturbative effects place an upper bound on the radius of convergence of the classical expansion and offer a potential resolution of the problem. In this paper, we elaborate more on these questions in the simpler, and slightly different, setting of wave-particle scattering. This has the advantage of giving a clear and intuitive picture with the benefit that all-orders analytical results are available in the literature. In particular, we will show that our classical radiative observables also generically exhibit mass singularities in the high-energy limit, analogous to those found in the classical two-body problem, which are completely resolved \emph{not} by higher-order classical effects but only within a quantum approach.

We consider in this paper the scattering of a massive scalar field on gauge and gravitational plane-wave backgrounds. Such systems have proven to be useful playgrounds for the study of scattering observables due to the high degree of symmetry of plane waves. Working in background field perturbation theory, arbitrarily strong plane waves can be treated analytically and exactly, with high-order scattering amplitudes on the background now available; see~\cite{Fedotov:2022ely} for a review. The recent literature even offers all-multiplicity results~\cite{Adamo:2020syc} and all-loop resummation of observables~\cite{Torgrimsson:2021wcj,Torgrimsson:2022ndq}. The simplicity of many of these results allows one to study both the quantum and classical regimes of wave scattering observables in detail. Furthermore, results can be re-expanded in powers of the background, effectively treating it as weak (which we will demonstrate is enough for our purposes), and then related to familiar objects in standard perturbation theory in vacuum.

This paper is organized as follows. We present our conventions below. In Section 2 we describe our wave-particle scattering
setup in both electrodynamics and gravity, along with the on-shell observables of interest. We show in particular that the angular momentum flow is a good local observable, by relating it to the phase and amplitude of the waveform.  The evaluation of our observables in electrodynamics is performed in Section 3, as part of which we highlight the role played by the Compton scattering cross section. This leads us to confront, in Section 4, the corresponding gravitational Compton cross section and its collinear divergence. Using the KLN theorem we show that the divergence is cancelled by forward scattering contributions, generalising the gauge theory results of~\cite{Frye:2018xjj}. This motivates our proposal to dress the asymptotic states to address collinear divergences in gravitational wave-particle scattering. Using the simplest choice of such a dressing, we proceed to the evaluation of our on-shell gravitational observables in Section 5. In Section 6 we analyse the high-energy behaviour of our observables. Classically, these generically exhibit mass singularities in the high energy limit. We show that these singularities are only resolved by quantum effects. We conclude in Section 7.

\paragraph{Conventions.} We set $c=1$ and work in $d=4$ unless stated explicitly, using lightfront coordinates $\ud s^2 = 2 \ud x^\LCp \ud x^\LCm - \ud x^a \ud x^a$ where  $x^{\pm} = \frac{1}{\sqrt{2}} (x^0 \pm x^3)$ and $a\in \{1,2\}$ spans the spatial `transverse' directions. We introduce two null vectors $n_\mu$, $\ell_\mu$ obeying $n\cdot \ell =1$, $n\cdot x = x^\LCm$ (`lightfront time') and $\ell\cdot x = x^\LCp$. We define $\hat{\delta}^n(\cdot) := (2 \pi)^n \delta(\cdot)$ and $\hat{\ud}^n q := \ud^n q /(2 \pi)^n$. In our conventions, $e$ is the classical electromagnetic charge, $\alpha =e^2/(4\pi\hbar)$ is the fine structure constant, $\kappa$ is the gravitational coupling and $G=\kappa^2/(32\pi)$ is Newton's constant.

\section{Wave particle scattering in QFT}
\label{sec:wavescattering}

We will study the quantum and classical scattering problem for a minimally coupled massive scalar field in a gauge or a gravitational plane-wave background. To describe the scattering process we prepare an incoming state $\ket{\text{in}}$ as the superposition
\begin{align}
  \ket{\text{in}} = \ket{\psi}\otimes\ket{\beta}
   \label{eq:instate}
\end{align}
of a wavepacket $\ket{\psi}$ for the massive scalar {$\phi$ of mass $m$} and a coherent state of photons or gravitons~$\ket{\beta}$.
In quantum electrodynamics (QED), we use the standard on-shell mode expansion for the scalar $\phi$ and the photon field $\mathbb{A}_{\mu}$
\begin{align}
\mathbb{\phi}(x)&=\frac{1}{\sqrt{\hbar}} \int\!\ud \Phi(p)\left[a(p) e^{-i \frac{p \cdot x}{\hbar}}+ b^{\dagger}(p) e^{i \frac{p \cdot x}{\hbar}}\right]\,,
\label{eq:onshell-oper-scalar} \\
\mathbb{A}_{\mu}(x)&=\frac{1}{\sqrt{\hbar}} \sum_{\sigma=\pm} \int\!\ud \Phi(k)\left[a_{\sigma}(k) \varepsilon_{\mu}^{\sigma *}(k) e^{-i \frac{k \cdot x}{\hbar}}+ a^{\dagger}_{\sigma}(k) \varepsilon_{\mu}^{\sigma}(k) e^{i \frac{k \cdot x}{\hbar}}\right] \;,
\label{eq:onshell-oper-photon}
\end{align}
where $\varepsilon_\mu^\sigma(k)$ are photon polarisation vectors of definite helicity $\sigma= \pm 1$, and the Lorentz-invariant phase space measure is
\begin{align}
     \mathrm{d} \Phi(k) \equiv \hat{\mathrm{d}}^{4} k \, \hat{\delta}\left(k^{2}\right) \theta\left(k^{0}\right) \,, \qquad   
    \mathrm{d} \Phi(p) \equiv \hat{\mathrm{d}}^{4} p \, \hat{\delta}(p^{2}-m^2) \theta(p^{0}) \,.
\end{align}
Turning to gravity, we treat general relativity (GR) as an effective field theory valid below the Planck scale~\cite{Donoghue:1994dn}, and expand the metric in perturbation theory as
\be
g_{\mu \nu} = \eta_{\mu \nu} + \kappa \, h_{\mu \nu} \,,
\ee
where the linearized graviton perturbation $h_{\mu \nu}$ corresponds to the quantum operator $\mathbb{h}_{\mu\nu}$,~i.e.
\be
\mathbb{h}_{\mu \nu}(x) =\frac{1}{\sqrt{\hbar}} \sum_{\sigma=\pm} \int\!\ud\Phi(k)\left[a_{\sigma}(k) \varepsilon_{\mu \nu}^{\sigma *}(k) e^{-i \frac{k \cdot x}{\hbar}}+ a^{\dagger}_{\sigma}(k)  \varepsilon_{\mu\nu}^{\sigma}(k) e^{i \frac{k \cdot x}{\hbar}}\right] \,,
\label{eq:onshell-oper-graviton}
\ee
with the graviton polarisation tensor $\varepsilon_{\mu\nu}^\sigma(k)$ of helicity $\sigma = \pm 2$. 

Using \eqref{eq:onshell-oper-scalar} we can write the incoming scalar state as
\begin{align}
\ket{\psi} &:= \int\!\ud \Phi(p) \phi(p)  e^{i \frac{p \cdot b}{\hbar}} a^{\dagger}(p) \ket{0}\,,
\end{align}
where $\phi(p)$ is a wavepacket and $b^{\mu}$ is a shift of this wavepacket in position space -- we make this precise later. The incoming photon or graviton coherent state is defined as
\be
\ket{\beta} := \mathcal{N}_{\beta} \exp\bigg( \sum_\sigma\int \!\ud \Phi(k) \beta^{\sigma}(k) a^{\dagger}_{\sigma}(k)\bigg) \ket{0} \,,
\label{eq:coherent-GR-pw}
\ee
with $a_\sigma^\dagger(k)$ the appropriate creation operator, $\beta^{\sigma}(k)$ the coherent state profile, or waveshape, and $\mathcal{N}_\beta$ a normalisation constant. We will choose specific waveshapes in a moment.

\begin{figure}[t!]
\centering
\includegraphics[scale=1.2]{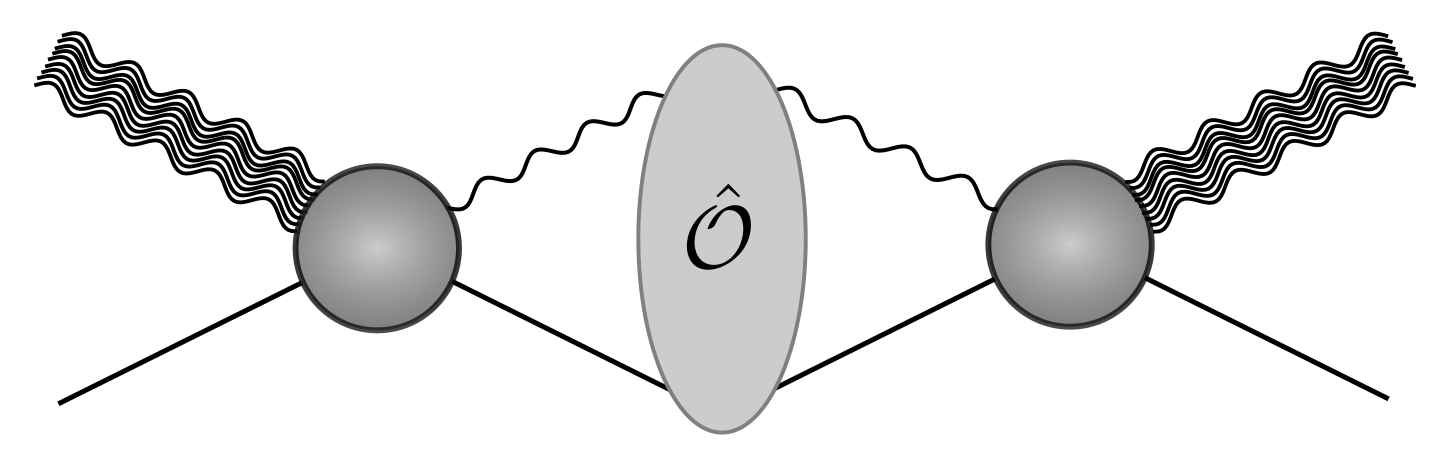}
\caption{The leading contribution to the expectation value of the operator $\hat{\mathcal{O}}$ for the radiation emitted by a scalar in a plane-wave background comes from the Compton amplitude.}
\label{fig:observable}
\end{figure}

We are interested in the radiative observables of this scattering problem. Observables are computed from expectation values of operators $\hat{O}$ in the final state $\ket{\text{out}}:= S \ket{\text{in}}$ defined by evolution with the S-matrix $S$;
\begin{align}
\langle \hat{O} \rangle := \bra{\text{out}} \hat{O} \ket{\text{out}} = \bra{\text{in}} S^{\dagger} \hat{O} S \ket{\text{in}}\,,
\label{eq:expect-operator}
\end{align}
as illustrated in Fig.~\ref{fig:observable}. We are in particular interested in analyzing the $\hbar$ expansion of the expectation value $\eqref{eq:expect-operator}$ as $\hbar \to 0$ in order to understand the properties of quantum and classical contributions~\cite{Iwasaki:1971vb,Krivitsky:1991vt,Higuchi:2002qc,Holstein:2004dn,Ilderton:2013tb,Kosower:2018adc}.
In practice, such expectation values are evaluated by inserting complete sets of asymptotic states to either side of the operator, which immediately yields an expression for the expectation value in terms of scattering amplitudes. The basis of such states is arbitrary; a normal number state basis is natural for perturbative calculations in vacuum, for example. We use instead a basis of  ``displaced number states'', defined by
\be
    1\!\!1 = \mathbb{D}(\beta)\mathbb{D}^\dagger(\beta) = \mathbb{D}(\beta)\sum\limits_n \ket{n}\bra{n} \mathbb{D}^\dagger(\beta) \;,
\ee
in which $\mathbb{D}(\beta)$ is the displacement operator which creates coherent states\footnote{Some of the key properties of  $\mathbb{D}(\beta)$, like unitarity, are reviewed in \cite{Ilderton:2017xbj,Cristofoli:2021vyo}.}, while the sum is, in compact notation, over number states of scalars, photons and gravitons in vacuum. This expresses our observables in terms of scattering amplitudes on a \emph{background} metric or gauge field, which we will have use of below. To illustrate this, we turn to the backgrounds of interest and explain how they are related to the coherent scattering states.

\subsection{Classical backgrounds from coherent states}
Consider the expectation value of the gauge potential $\mathbb{A}_\mu$ in QED, evaluated in the time-evolved coherent state~\eqref{eq:coherent-GR-pw}:
\begin{align}
\label{eq:expecQED}
\bra{\beta} S^{\dagger} \mathbb{A}_{\mu}(x) S \ket{\beta}  &:= \bra{0} \mathbb{D}^{\dagger}(\beta)\, S^{\dagger} \mathbb{A}_{\mu}(x) S \, \mathbb{D}(\beta)\ket{0} \\
&= \bra{0} \, S^{\dagger} [A_{\text{cl}}] (\mathbb{A}_{\mu}(x) + A_{\mu}^{\text{cl}}(x))  S [A_{\text{cl}}] \ket{0} \nonumber \\
&=  \bra{0} \, S^{\dagger} [A_{\text{cl}}]  \mathbb{A}_{\mu}(x) S [A_{\text{cl}}] \ket{0} + A_{\mu}^{\text{cl}}(x) \,, \nonumber 
\end{align}
in which we have used the key property that the displacement operator translates the creation/annihilation operators by, essentially, the waveshape $\beta(k)$; it therefore translates the field operator by a classical field $A_{\text{cl}}^\mu$ which, from (\ref{eq:onshell-oper-photon}), is the on-shell Fourier transform of the waveshape. As such the $S$-matrix on the classical background $A^\mu_{\text{cl}}$ appears:
\be
    S[A_{\text{cl}}]  := {\mathbb{D}^{\dagger}(\beta_{\text{QED}}) S \mathbb{D}(\beta_{\text{QED}})\,.}
\ee
The first term in the final line of (\ref{eq:expecQED}) is the one-point function $\langle \mathbb{A}_{\mu}(x) \rangle_{A_{\text{cl}}}$ on the background. It is nonzero for general $\beta$ but, because QED is an abelian theory, is purely quantum mechanical, the first contribution being the one-loop `tadpole' in the background~$A_{\text{cl}}$~\cite{Gies:2016yaa,Ahmadiniaz:2017rrk,Ahmadiniaz:2019nhk,Karbstein:2019wmj}. It is moreover zero for the $A^\mu_{\text{cl}}$ we will consider. We therefore have, in the classical limit,
\be
\bra{\beta} S^{\dagger} \mathbb{A}_{\mu}(x) S \ket{\beta} \Big|_{\hbar \to 0} =  A_{\mu}^{\text{cl}}(x) \;. 
\label{eq:expecQED-final}
\ee
It is time to specify the waveshape, equivalently the classical field, of interest: this is a plane wave, conveniently represented by the potential
\be\label{eq:Amu-position-QED}
A^\mu_{\text{cl}}(x) := - n_\mu x^a E_a(x^\LCm) \;,
\ee
in which the two real functions $E_a(x^\LCm)$ parameterise the physical electromagnetic fields of the wave. We take these to be compactly supported, say $x^\LCm \in [0,T]$ with $T>0$, which separates the spacetime into well-defined asymptotic regions. We will not need the explicit expression for $\beta(k)$, for a discussion of which see~\cite{Cristofoli:2022phh}.

Turning to gravity, the waveshape is chosen to describe a gravitational plane wave. In Brinkmann coordinates this is given by~\cite{Brinkmann:1925fr}
\be\label{eq:Brinkmann}
    h^{\text{cl}}_{\mu\nu}(x) = - x^a x^b H_{ab}(x^\LCm) n_\mu n_\nu \;,
\ee
where the vacuum equations require ${H}_{ab}$ to be a symmetric, traceless $2\times2$ matrix. We take this to be compactly supported in $x^\LCm$, as we did in QED. The relation to coherent scattering states is, though, more subtle than in QED. The expectation value of the graviton field $h_{\mu \nu}(x)$ between time-evolved coherent states of gravitons is, proceeding as in (\ref{eq:expecQED}),
\begin{align}
\bra{\beta} S^{\dagger} \mathbb{h}_{\mu \nu}(x) S \ket{\beta}
&=  h_{\mu \nu}^{\text{cl}}(x)  + \bra{0} \, S^{\dagger}[h_{\text{cl}}] \mathbb{h}_{\mu \nu}(x) S[h_{\text{cl}}]\ket{0} \,,
\label{eq:expec_GR}
\end{align}
and the S-matrix on the background $h_{\mu \nu}^{\text{cl}}(x)$ is now $S[h_{\text{cl}}]  := \mathbb{D}^{\dagger}(\beta_{\text{GR}}) S \mathbb{D}(\beta_{\text{GR}})$.

\begin{figure}[!t]
\centering
\includegraphics[scale=1.1]{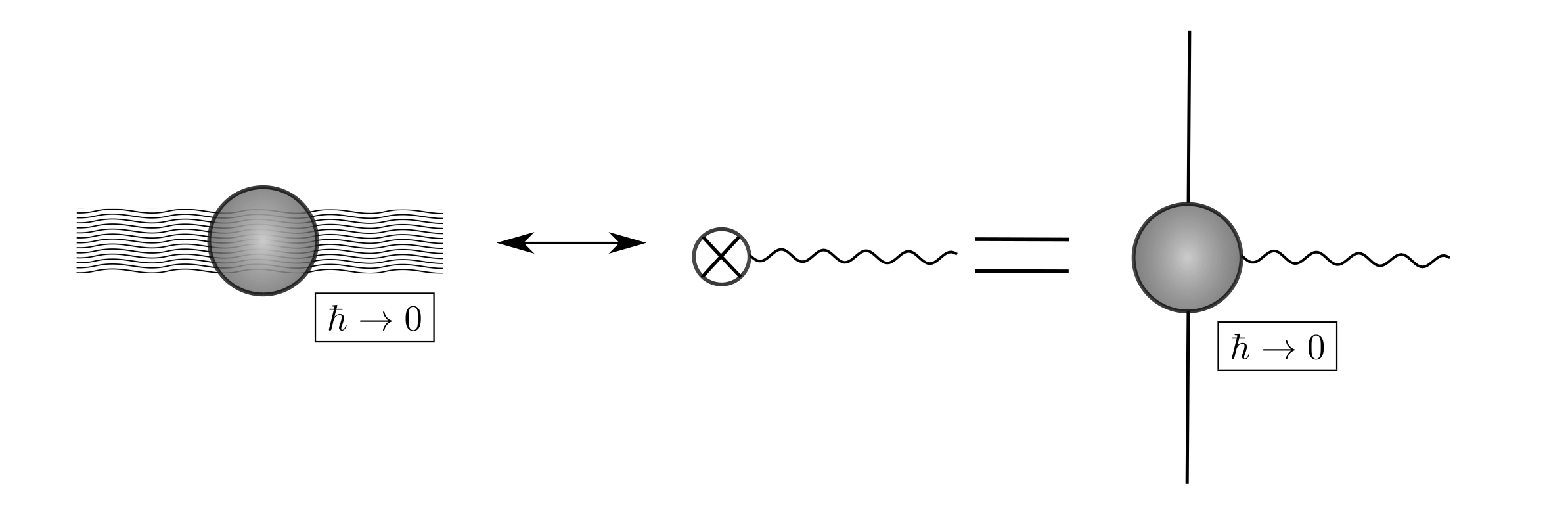}
\caption{Comparison between, left, vacuum solutions such as plane-waves and, right, point-like backgrounds. Both are generated in the classical limit $\hbar\to0$ by time-evolution (indicated by the `blob') from an initial state; for the plane wave this is a coherent state, while for the point-like background it is a particle coherently emitting photons/gravitons.
}
\label{fig:backgrounds}
\end{figure}

Due to the self-interaction of the gravitational field, the one-point function $\langle \mathbb{h}_{\mu \nu}(x) \rangle_{h_{\text{cl}}}$ in a background is generally non-vanishing, even classically. Here the symmetries of the gravitational plane-wave come into play: for this geometry there is no particle production~\cite{Gibbons:1975jb}, as can be shown by a direct calculation~\cite{Adamo:2017nia}. We therefore obtain, as desired, 
\begin{align}
\bra{\beta} S^{\dagger} \mathbb{h}_{\mu \nu}(x) S \ket{\beta} \Big|_{\hbar \to 0}  &=- x^a x^b H_{ab}(x^\LCm) n_\mu n_\nu \,.
\end{align}
Plane waves are, in both gravity and electromagnetism, highly symmetric vacuum solutions with five isometries in $d=4$, corresponding to invariance under three translations and two boosts in null directions~\cite{Baldwin,Ehlers,Duval:2017els}. This high degree of symmetry will allow us to give compact and revealing expressions for scattering observables. Let us compare briefly with point-like backgrounds, which arise from the S-matrix time evolution of external particles coupled to gauge or gravitational fields in the classical limit~\cite{Monteiro:2020plf}, see Fig.~\ref{fig:backgrounds}. Usually, these calculations require one to consider the perturbative evaluation of the three-point function with two external particles and one messenger, which has to be considered off-shell in Lorentzian signature. In particular, in QED and GR we define point-like backgrounds as
\begin{align}
A_{\mu}^{\text{cl}} := \bra{p} S^{\dagger} \mathbb{A}_{\mu}(x) S \ket{p} \Big|_{\hbar \to 0} \,,  \qquad h_{\mu \nu}^{\text{cl}} &:= \bra{p} S^{\dagger} \mathbb{h}_{\mu \nu}(x) S \ket{p} \Big|_{\hbar \to 0}\,,  
\end{align}
where $\ket{p}$ is the external on-shell scalar state and the domain of the field operators $\mathbb{A}_{\mu}(x)$ and $\mathbb{h}_{\mu \nu}(x)$ is extended to include off-shell contributions, as in the standard Schwinger-Keldysh formalism.  At any order in perturbation theory, this defines an effective classical source which generates the gauge potential or metric in the classical theory via time evolution: relevant solutions like Coulomb, Schwarzschild and shockwave metrics are all generated by this procedure, see e.g.~\cite{Duff:1973zz,Moynihan:2019bor,Cristofoli:2020hnk,Guevara:2021yud}. This provides an alternative way to generate a relevant class of gauge potentials or classical metrics in gravity, directly from scattering amplitudes with external particles or waves~\cite{Neill:2013wsa,Bjerrum-Bohr:2018xdl,Monteiro:2020plf}.

\subsection{On-shell wave observables: from global to local}

We introduce now the set of global and local observables for wave-particle scattering that we will later compute in Sections \ref{sec:QEDobservables} and \ref{sec:GR-obs-section}.  Natural global observables are the scattering cross-section, or scattering probability. Other global observables are the total momentum and angular momentum, in particular their change due to a scattering process. The total radiated momentum takes the same form in electromagnetism and gravity, being the expectation value of the operator 
\be\label{eq:K-both}
\mathbb{K}^{\mu} := \sum_{\sigma = \pm } \int \!\ud \Phi(k) \, k^{\mu} a^{\dagger}_{\sigma} (k) a_{\sigma}(k) \;, 
\ee
for the appropriate choice of photon or graviton mode operators.

A gauge-invariant expression for the total angular momentum was recently provided by~\cite{Manohar:2022dea,DiVecchia:2022owy}; in QED the angular momentum operator for the gauge field is\footnote{Our conventions are that $A_{[\mu}B_{\nu]}=A_{\mu} B_{\nu}-A_{\nu} B_{\mu}$, while $\overleftrightarrow{\partial} = \frac{1}{2} (\overrightarrow{\partial} - \overleftarrow{\partial} )$.}
\be
\label{eq:KJ-QED}
\mathbb{J}^{\mu \nu}_{\text{QED}} := \sum_{\sigma = \pm } \int\!\ud \Phi(k) \varepsilon^{\alpha}_{\sigma}(k) a^{\dagger}_{\sigma}(k)\left[(\mathcal{J}_{\text{QED}})^{\mu \nu}_{\alpha \beta} \right] \varepsilon^{*\beta}_{\sigma}(k) a_{\sigma}(k) \,, 
\ee
where the mode operators $a_{\sigma}(k)$ and $a^{\dagger}_{\sigma}(k)$ are regarded as functions of $k^{\mu}$ and
\begin{align}
(\mathcal{J}_{\text{QED}})^{\mu \nu}_{\alpha \beta} &= -i \eta_{\alpha \beta} k^{[\mu} \frac{\stackrel{\leftrightarrow}{\partial}}{\partial k_{\nu]}}  -i \delta^{[\mu}_{\alpha} \delta^{\nu]}_{\beta} \,.
\end{align}
For gravitons the relevant operator is similar, up to polarisation structure: \begin{align}
\label{eq:KJ-GR}
\mathbb{J}^{\mu \nu}_{\text{GR}} &= \sum_{\sigma = \pm } \int\!\ud \Phi(k) \varepsilon^{\alpha \alpha'}_{\sigma}(k) a^{\dagger}_{\sigma}(k)\left[(\mathcal{J}_{\text{GR}})^{\mu \nu}_{\alpha \alpha' \beta \beta'}
\right] \varepsilon^{*\beta \beta'}_{\sigma}(k) a_{\sigma}(k) \,,
\end{align}
in which
\begin{align}
(\mathcal{J}_{\text{GR}})^{\mu \nu}_{\alpha \alpha' \beta \beta'} &=+ i\eta_{\alpha \beta} \eta_{\alpha' \beta'} k^{[\mu} \frac{\stackrel{\leftrightarrow}{\partial}}{\partial k_{\nu]}}  
+2 i \eta_{\alpha' \beta'} \delta^{[\mu}_{\alpha} \delta^{\nu]}_{\beta} \,.
\end{align}
We turn to local observables. Just as we can consider, e.g.~the differential rather than total cross-section, we can construct local observables as the differential analogues of the global observables above, when they are well-defined. For example, a detector placed in a particular direction $\hat{v}$ on the celestial sphere gives access to the angular dependence of observables related to outgoing particles and radiation. The `waveform' is a common example ~\cite{Cristofoli:2021vyo}: it is a proper local observable which can be computed as the leading on-shell component of the radiation field at large distances.

The local analogues of the total momentum and angular momentum are the momentum and angular momentum `flow', recently discussed for scalar, vector and tensor radiative fields in~\cite{Gonzo:2020xza}. These are trivially obtained from (\ref{eq:K-both})--\eqref{eq:KJ-GR} by inserting a delta-function under the integrals which selects out the direction of interest. Thus the momentum flow, which we denote by $\mathcalbb{P}^{\mu}$, is 
\be
    \mathcalbb{P}^{\mu}= \sum_{\sigma = \pm } \int\! \ud \Phi(k) \, \hat{\delta}^2(\Omega - \Omega_{\hat{v}}) \, k^{\mu} a^{\dagger}_{\sigma} (k) a_{\sigma}(k) \,,
\ee
with, as above, the appropriate choice of mode operator for electrodynamics or gravity, and $\Omega$ denotes the angular measure. Here we have introduced the convenient notation $\hat\delta^2(\Omega-\Omega_{\hat v})=(2\pi)^2\delta^2(\Omega-\Omega_{\hat v})$. The momentum flow has been connected to the \emph{amplitude} of the waveform in~\cite{Cristofoli:2021vyo}, confirming that $\mathcalbb{P}^\mu$ is a well-defined, IR finite, observable.

This naturally prompts the question: what about the \emph{phase} of the waveform? To address this, we introduce the local analogue of $\mathbb{J}^{\mu\nu}$, the angular momentum \emph{flow} $\mathcalbb{N}^{\mu \nu}$:
\begin{align}
\label{eq:light-ray_flowQED}
\mathcalbb{N}^{\mu \nu}_{\text{QED}} &= \sum_{\sigma = \pm } \int\!\ud \Phi(k) \, \hat{\delta}^2(\Omega - \Omega_{\hat{v}}) \, \varepsilon^{\alpha}_{\sigma}(k) a^{\dagger}_{\sigma}(k) \left[(\mathcal{J}_{\text{QED}})^{\mu \nu}_{\alpha \beta} \right] \varepsilon^{*\beta}_{\sigma}(k) a_{\sigma}(k) \,, \\
\label{eq:light-ray_flowGR}
\mathcalbb{N}^{\mu \nu}_{\text{GR}} &=  \sum_{\sigma = \pm } \int\!\ud \Phi(k)\, \hat{\delta}^2(\Omega - \Omega_{\hat{v}}) \,  \varepsilon^{\alpha \alpha'}_{\sigma}(k) a^{\dagger}_{\sigma}(k) \left[(\mathcal{J}_{\text{GR}})^{\mu \nu}_{\alpha \alpha' \beta \beta'}  \right] \varepsilon^{*\beta \beta'}_{\sigma}(k) a_{\sigma}(k)  \,.
\end{align}
We will now give a new and direct connection between the angular momentum flow and the phase of the waveform. We make this connection in the classical limit, where on general grounds one expects the outgoing radiation created in a scattering process to be described by a coherent state\footnote{The general quantum case can be treated by making use of a superposition of coherent states, and the arguments in this section can be generalised to that case using the Glauber-Sudarshan P-representation \cite{Sudarshan:1963ts,Glauber:1963tx}, see~\cite{Ekman:2020vsc} for related comments in the context of back-reaction.}~\cite{Cristofoli:2021jas,Britto:2021pud,DiVecchia:2022piu}, except possibly for static zero-energy contributions \cite{DiVecchia:2022piu,Herderschee:2023fxh,Georgoudis:2023lgf,Brandhuber:2023hhy}. Let the corresponding wave-shape for this state be $\BETA^{\sigma}$ to distinguish it from our incoming state $\beta^\sigma$. Beginning in QED, we consider the coherent state 
\begin{align}
    \ket{\BETA}_{\text{QED}} = \mathcal{N}_{\BETA} \exp\left(\sum_{\sigma} \int \!\ud \Phi(k) \, \BETA^{\sigma}(\omega, \omega \hat{k}) a^{\dagger}_{\sigma}(k)\right) \ket{0} \,,
\end{align}
and split the waveform into a basis $\{v^{\mu}$,$\bar{v}^{\mu},m^{\mu}$,$\bar{m}^{\mu}\}$ where $v^{\mu}$,$\bar{v}^{\mu}$ are null, $v^{\mu} = (1,\hat{v})$ with $\hat{v}$ the direction in which we measure the local observable, and $m^{\mu}$,$\bar{m}^{\mu}$ are spacelike, corresponding to a basis of polarisation vectors. We then identify two scalar projections of the waveform in this basis, and decompose them into real amplitudes $A_{\pm}$ and phases $\delta_{\pm}$ as
\begin{align}\label{eq:A-delta-def}
    &A_{\pm}(\omega, \omega \hat{v}) e^{i \delta_{\pm}(\omega, \omega \hat{v})} := m_{\pm}^{\mu} \varepsilon^{*(\pm)}_{\mu}(\hat{v}) \BETA^{(\pm)}(\omega, \omega \hat{v})\,.
\end{align}
Here $m_+^\mu=m^\mu$ and $m_-^\mu=\bar m^\mu$, with $m_\sigma^\mu\varepsilon^{*(\sigma)}_\mu(\hat v)=-1$.

In terms of these, the expectation value of the linear momentum flow is indeed expressed in terms of the amplitude~\cite{Cristofoli:2021vyo},
\begin{align}
    \bra{\BETA} \mathcalbb{P}^{\mu} \ket{\BETA}_{\text{QED}} &= \sum_{\sigma = \pm} \int \!\ud \Phi(k) \hat{\delta}^{2}(\Omega - \Omega_{\hat{v}}) \, k^{\mu} |A_{\sigma}(\omega, \omega \hat{v})|^2  \\
     &= \sum_{\sigma = \pm} \int \frac{\!\ud \omega \, \omega}{4 \pi} \, \omega v^{\mu} |A_{\sigma}(\omega, \omega \hat{v})|^2 \;, \nonumber
\end{align}
while the angular momentum flow becomes
\begin{align}
    \bra{\BETA} \mathcalbb{N}^{\mu \nu} \ket{\BETA}_{\text{QED}} &= \sum_{\sigma = \pm } \int\!\ud \Phi(k) \hat{\delta}^{2}(\Omega - \Omega_{\hat{v}}) \varepsilon^{\alpha}_{\sigma}(k) \BETA^{*\sigma}(k) \left[(\mathcal{J}_{\text{QED}})^{\mu \nu}_{\alpha \beta} \right] \varepsilon^{*\beta}_{\sigma}(k) \BETA^{\sigma}(k)   \\
    &= \sum_{\sigma = \pm }  \int \frac{\!\ud \omega \, \omega}{4 \pi} |A_{\sigma}(\omega, \omega \hat{v})|^2 \, \left[ - k^{[\mu} \frac{\partial \delta_{\sigma}(k)}{\partial k_{\nu]}} -i \varepsilon_{\sigma}^{[\mu}(\hat{k}) \varepsilon_{\sigma}^{*\nu]}(\hat{k})\right] \Bigg|_{\hat{k} =\hat{v}}\,. \nonumber
\end{align}
This shows that, at least in the classical limit, knowledge of the amplitude and the phase of the waveform is enough to completely determine the momentum and angular momentum flows, and it provides a new connection between these observables on the celestial sphere. We simply state the corresponding expression for gravity in terms of the analogue of (\ref{eq:A-delta-def}):
\begin{align}
    \hspace{-10pt}\bra{\BETA} \mathcalbb{N}^{\mu \nu}\ket{\BETA}_{\text{GR}}  = \sum_{\sigma = \pm }  \int \frac{\!\ud \omega \, \omega}{4 \pi} |A_{\sigma}(\omega, \omega \hat{v})|^2 \, \left[ -k^{[\mu} \frac{\partial \delta_{\sigma}(k)}{\partial k_{\nu]}} +2 i \eta_{\alpha \beta} \varepsilon_{\sigma}^{\alpha [\mu}(\hat{k}) \varepsilon_{\sigma}^{*\nu] \beta}(\hat{k})\right] \Bigg|_{\hat{k} =\hat{v}}\,.
\end{align}
We conclude this section with a comment on the gauge invariance {of local observables}. The linear momentum flow is trivially gauge-invariant since it does not explicitly depend on the polarisation vectors, while the angular momentum flow requires some work. For QED, a gauge transformation $\varepsilon^{\mu}_{\sigma}(k) \to \varepsilon^{\mu}_{\sigma}(k) + \xi k^{\mu}$ generates the following additional contribution to the angular momentum flow:
\begin{align}\label{change}
    \Delta_{\xi} \mathcalbb{N}^{\mu \nu}_{\text{QED}} &= \frac{1}{2} \xi\sum_{\sigma = \pm } \int\!\ud \Phi(k) \, \hat{\delta}^2(\Omega-\Omega_{\hat{v}}) a^{\dagger}_{\sigma}(k)\left[k_{\alpha}  k^{[\mu} \frac{\partial \varepsilon_{\sigma}^{*\alpha}(k)}{\partial k_{\nu]}}  +  k^{[\mu} \varepsilon_{\sigma}^{*\nu]}(k)\right] a_{\sigma}(k) \\
    &+ \frac{1}{2} \xi\sum_{\sigma = \pm } \int\!\ud \Phi(k) \, \hat{\delta}^2(\Omega-\Omega_{\hat{v}}) a^{\dagger}_{\sigma}(k)\left[k_{\alpha}  k^{[\mu} \frac{\partial \varepsilon_{\sigma}^{\alpha}(k)}{\partial k_{\nu]}}  +  \varepsilon_{\sigma}^{[\mu}(k) k^{\nu]}\right] a_{\sigma}(k) \;.  \nonumber
\end{align}
One might suspect that integration-by-parts would kill these terms, but we cannot use this, as the integral is implicitly absent, being removed by the delta-function. Instead we observe that transversality of the polarisation vectors, $k\cdot\varepsilon_\sigma(k)=0$ implies the useful identity
\begin{align}
 \frac{\partial}{\partial k_\mu} (k\cdot \varepsilon_\sigma(k))= 0 \implies k_\alpha \frac{\partial \varepsilon_\sigma^\alpha(k)}{\partial k_\mu}  = -\varepsilon_\sigma^\mu(k) \;.
\end{align}
Inserting this into (\ref{change}) we immediately find, as desired, that $\Delta_{\xi} \mathcalbb{N}^{\mu \nu}_{\text{QED}} = 0$. A straightforward calculation similarly confirms the gauge invariance of $\mathcalbb{N}^{\mu \nu}_{\text{GR}}$.

Table~\ref{tab:observables} summarises the observables to be considered.
\begin{table}[t!!]
    \makegapedcells
  \centering
\begin{tabular}{c|c|c}  \toprule 
    {\text{Theory}} & {\text{Global radiative observables}} & {\text{Local radiative observables}}  \\ \midrule
    \text{QED}  & $\sigma, \mathbb{K}^{\mu}_{\text{QED}},\mathbb{J}^{\mu \nu}_{\text{QED}}$  & $\frac{d \sigma}{d \Omega}, \mathcalbb{P}^{\mu}_{\text{QED}},\mathcalbb{N}^{\mu \nu}_{\text{QED}}$  \\ \midrule
    \text{GR}  & $\sigma, \mathbb{K}^{\mu}_{\text{GR}},\mathbb{J}^{\mu \nu}_{\text{GR}}$  & $\frac{d \sigma}{d \Omega}, \mathcalbb{P}^{\mu}_{\text{GR}},\mathcalbb{N}^{\mu \nu}_{\text{GR}}$  \\ \bottomrule
\end{tabular}
\caption{Summary of the on-shell global and local radiative observables of interest.}
\label{tab:observables}
\end{table}

\section{Wave scattering observables in quantum electrodynamics}
\label{sec:QEDobservables}
\begin{figure}[!t]
\centering
\includegraphics[scale=0.8]{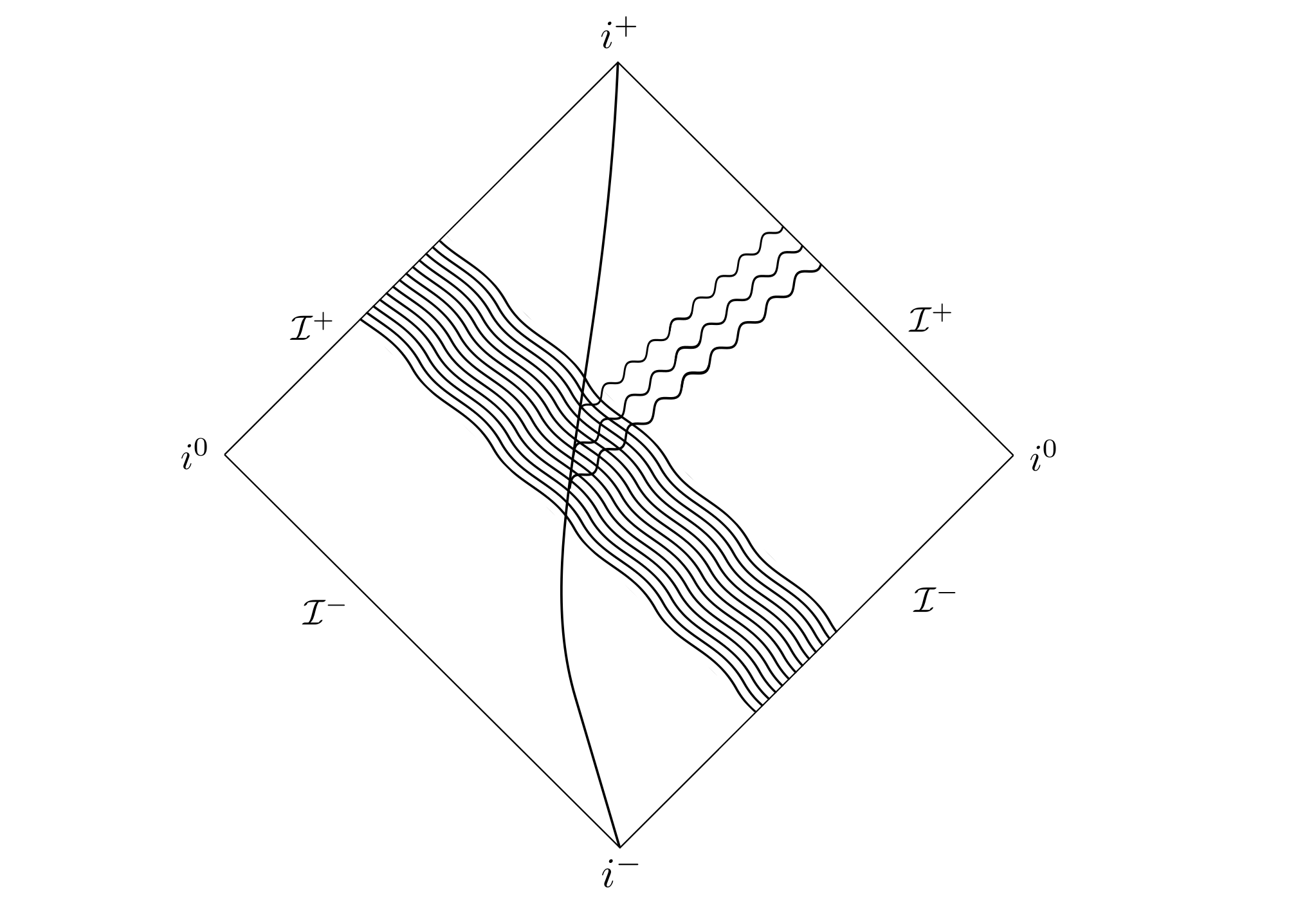}
\caption{We study the quantum and classical radiative observables for the wave emitted during the interaction of the massive scalar field, identified with the trajectory from $i^-$ to $i^+$, and the plane-wave background represented by the tightly-spaced wiggly lines.}
\label{fig:scattering}
\end{figure}

The scattering process of interest is illustrated in Fig.~\ref{fig:scattering}: the massive field enters (resp. leaves) the wave at the lightfront time $n\cdot x=0$ (resp. $n\cdot x = T$), emitting radiation whose properties we wish to study. We define the scalar wavepacket such that $b_\mu$ is its center at lightfront time $n\cdot x=0$; in the classical limit, this will essentially become the position of the scalar when it hits the wave\footnote{As such, $b_\mu$ does not have the same interpretation as the impact parameter in classical $2\to2$ scattering~\cite{Kosower:2018adc}.}. In this section, we will compute the wave scattering observables discussed in section \ref{sec:wavescattering} in quantum (and classical) electrodynamics at leading order in the perturbative expansion.

Before doing so, it is worth discussing the relevant length scales in the problem. The first is given by the electron radius $r_e = e^2 / (4 \pi m)$, which controls the weak coupling expansion. The natural length scale of quantum effects is the Compton wavelength $\lambdabar_C = \hbar/m$. A third, classical, length scale, emerges from the observation that, for a field of typical wavenumber $\bar{k}=\bar{\omega} n$ (localised along the $n_\mu$ direction),  $\bar{\omega}\,  n\cdot p = {\bar k}\cdot p$ is the invariant product of particle energies.
Hence $\bar{\omega}\, n\cdot p/m$ is the invariant wave frequency seen by the particle in its rest frame, and the corresponding classical length scale is $\lambdabar = m/ (\bar{\omega}\, n\cdot p)$. In contrast to the Compton wavelength, this goes like the mass, $\lambdabar\sim m$. Summarizing, we have 
\begin{align}
\label{eq:def_scalesQED}
r_e = \frac{e^2}{4 \pi m} \,, \qquad \lambdabar_C = \frac{\hbar}{m}\,, \qquad \lambdabar = \frac{m}{\bar{\omega}\, (n\cdot p)} \,.
\end{align}
For the calculation of quantum observables, we simply assume
\begin{align}
\label{eq:QED_quantum_observables}
r_e \ll \lambdabar_C,\lambdabar
\end{align}
while for classical observables we will impose the hierarchy
\begin{align}
\label{eq:CED_classical_observables}
r_e \ll \lambdabar_C \ll \lambdabar \,.
\end{align}

\subsection{Differential cross section}\label{subsec:cross-section-qed}
Natural observables in $2\to2$ and $1\to n$ scattering in vacuum are, respectively, the cross-section and decay rate.
A natural observable in $1\to n$ scattering on a plane wave background is the total scattering probability~\cite{Fedotov:2022ely}. We discuss this probability here, in particular its relation to the cross-section of Compton scattering in vacuum. Making this connection is important because the different behaviour of the cross-sections in QED and gravity has consequences (discussed in Section~\ref{sec:KLNtheorem}) for other observables.

The three-point amplitude $\mathcal{A}_3 = \bra{p';k^\sigma}S[A_\text{cl}]\ket{\psi}$ for photon emission on the plane wave background is~\cite{Ritus1985,Fedotov:2022ely}
\be\label{eq:amp-QED-OK}
	\mathcal{A}_3 =	\int\!\ud\Phi(p)\, \phi(p)e^{ib\cdot p}
	\hat{\delta}(n \cdot (p'+k - p)) \hat{\delta}^2_{\LCperp}(p'+k - p) \mathcal{I}_3(p,k^\sigma) \;,
\ee
in which the nontrivial part of the amplitude is, up to an irrelevant phase,
\be\begin{split}
\label{eq:amp-mess-qed}
    \mathcal{I}_3(p,k^\sigma) &=
    2e \int_0^T\!\ud x^\LCm\,  n\cdot (p-k)\exp\bigg[i\int^{x^\LCm}_0\!\ud s\, \frac{k\cdot P(s)}{n\cdot (p-k)}\bigg] \frac{\partial}{\partial x^\LCm} \left[\frac{\varepsilon_\sigma(k)\cdot P(x^\LCm)}{k\cdot P(x^\LCm)}\right] \;,
\end{split}
\ee
where the `dressed momentum' $P_\mu(x^\LCm)$ is
\be
    P_\mu(x^\LCm) = p_\mu - eA_\mu(x^\LCm) + n_\mu \frac{2eA(x^\LCm)\cdot p- e^2A^2(x^\LCm)}{2n\cdot p} \;, \quad  A_\mu(x^\LCm) := \int_{-\infty}^{x^\LCm}\!\ud y \, \delta_\mu^b E_b(y) \;.
\ee
The amplitude (\ref{eq:amp-mess-qed}) is manifestly all-orders in the coupling to the background $\sim eE$. The spacetime integral over $x^\LCm$ in (\ref{eq:amp-mess-qed}) cannot be performed exactly in general. Gauge invariance is manifest -- the integrand vanishes identically if $\varepsilon_\sigma(k)$ is replaced by $k$~\cite{Dinu:2012tj}. Mod-squaring and summing over final states yields the total scattering probability
\be\label{eq:QED-P-0}
\begin{split}
\mathbb{P}_{\text{QED}} = \int\!\ud\Phi(p)\, |\phi(p)|^2 \sum_\sigma\int\!\ud \Phi(k)\,
	\frac{1}{2 (n \cdot p)}
	\frac{ |\mathcal{I}_3(p,k^\sigma)|^2 }{2 n \cdot (p-k)}\;.
\end{split}
\ee
All dependence on $b_\mu$ drops out. Under the usual assumption that the wavepacket  is strongly peaked around a given momentum, it can be integrated out --  the effect is simply to set everything preceding the sum in (\ref{eq:QED-P-0}) to unity.

The phenomenology of the differential probability (essentially the emission spectrum) is very rich due to its all-orders nature~\cite{Fedotov:2022ely}, and has been probed in various experiments~\cite{E144:1996enr,Cole:2017zca,Poder:2017dpw}. Though we return to some all-orders results below, we restrict for the remainder of this section to leading-order perturbative results, in which the connection to the
Compton cross-section is buried. Expanding the amplitude~\eqref{eq:amp-QED-OK} in $e$, the tree-level contribution is
\begin{align}\label{eq:ampl-QED-leading}
\mathcal{A}_{3,\sigma}^{(0)} & = i e^2 \int\!\ud\Phi(p)\, \phi(p)e^{ib\cdot p}
	\hat{\delta}(n \cdot (p'+k - p)) \hat{\delta}^2_{\LCperp}(p'+k - p) \,  \\
&\qquad\qquad\times 2 \int_0^T\!\ud x^\LCm\,
e^{i \frac{k \cdot p}{n \cdot (p - k)} x^\LCm} \frac{(\varepsilon_{\sigma}(k) \cdot p)\, \sfx\, r^a -  (k \cdot p)\varepsilon^a_{\sigma}(k) }{k \cdot p} A_a(x^\LCm) \,, \nonumber
\end{align}
where we use the polarisation gauge $n\cdot\varepsilon_\sigma(k)=0$ and have parameterised the outgoing photon momentum by its lightfront momentum fraction $\sfx$, and a transverse two-vector $r_\LCperp$ defined by
\be\label{eq:defxr}
	\sfx:=\frac{n \cdot k}{n \cdot p} \;,
	\qquad
    r_\LCperp := \frac{k_\LCperp - \sfx p_\LCperp}{\sfx} \,.
\ee
Eq.~(\ref{eq:ampl-QED-leading}) has the interpretation of a \emph{Compton-scattering} amplitude, in which the incoming photon is directed along the plane-wave. To this order in perturbation theory, the background profile enters essentially as a wavepacket for this photon.

Squaring up, one finds that the (differential and) total probability (\ref{eq:QED-P-0}) takes the simplest form when parameterised in terms of the variables (\ref{eq:defxr}).  It will be important to understand the physical content of $r_\LCperp$. To do so, define the \emph{null} Lorentz boost
\be\label{theboost}
	\Lambda_{\mu\nu} = \exp \Big(\frac{1}{n \cdot p} n_{[\mu} \delta_{\nu]}^a p_a\Big) \;,
\ee
which, acting on \emph{any} on-shell $q_\mu$, shifts the transverse momenta as $q_\LCperp \to q_\LCperp - (n \cdot q / n \cdot p) p_\LCperp$, while leaving $n\cdot q$ unchanged. In particular, the boost kills the transverse momenta in $p_\mu$ and leaves the plane wave vector $n_\mu$ \emph{invariant}, thus (\ref{theboost}) takes us to the frame in which the wave-particle collision is head-on, and the photon has momentum degrees of freedom $n\cdot k$ and $k_\LCperp - \sfx p_\LCperp$, equivalently $\sfx$ and $r_\LCperp$. With our conventions the plane wave travels in the negative $z$-direction, so we define, along with an azimuthal angle $\phi$, a polar angle $\theta$ measured from the south pole, such that $\theta=0$ corresponds to propagation \emph{collinear} with the wave. One finds
\be\label{eq:angles}
r_\LCperp = \frac{\sqrt{2} (n \cdot p) \sin\theta}{1-\cos\theta} (\cos\phi,\sin\phi) 
\;,
\ee
showing that $r_\LCperp$ describes two angular degrees of freedom, with $\theta$ a boosted generalisation of the scattering angle in the `lab' frame, in which the collision would be head on.
The calculation from here is straightforward; the leading order scattering probability is
\be\label{eq:P2XQED}
    \mathbb{P}_{\text{QED}} \to  \int_0^\infty\!\frac{\ud\omega}{\pi \omega}\, |E_a(\omega)|^2 \, \int_{-1}^{1}\!\ud (\cos\theta)\, \frac{\pi \alpha^2 (1+\cos^2\theta)}{m^2(1+\nu (1-\cos \theta))^2} \;,
\ee
in which the dimensionless frequency $\nu$ is
\be\label{eq:frequency-nu}
    \nu \equiv \nu(\omega):= \frac{\omega n \cdot p}{m^2} 
\ee
and the Fourier transform of the electric field is
\begin{align}
    {E}_{a}(\omega) := \int_0^T\!\ud x^\LCm e^{i \omega x^\LCm} E_{a}(x^\LCm) \,.
\end{align}
We recognise the final factor in (\ref{eq:P2XQED}): it is the textbook differential Compton cross-section,
\be
    \frac{1}{2 \pi} \frac{\ud\sigma_{\text{QED}}(\nu,\theta)}{\ud \cos \theta}= \frac{\alpha^2 (1+\cos^2\theta)}{2 m^2(1+\nu (1-\cos \theta))^2} \,,
\ee
and so our (total or differential) scattering probability is given by the frequency convolution of the (total or angular-resolved) cross section with the wave profile:
\be
    \mathbb{P}_{\text{QED}} = \int_0^\infty\!\frac{\ud\omega}{\pi \omega}\, |E_a(\omega)|^2 \, \int_{-1}^{1}\!\ud (\cos\theta) \frac{\ud\sigma_{\text{QED}}(\nu,\theta)}{\ud \cos\theta} \;.
\ee

\subsection{The momentum flow $\mathcalbb{P}^{\mu}_{\text{QED}}$ and the total momentum $K^{\mu}_{\text{QED}}$}

The total radiated momentum produced in the wave scattering process is 
\begin{align}
    \langle \mathbb{K}^{\mu}\rangle_{\text{QED}}  &= \sum_\sigma\int\!\ud \Phi(k)\ud\Phi(p') k^{\mu} |\mathcal{A}_{3,\sigma}^{(0)}|^2\;.
\end{align}
Inserting the amplitude (\ref{eq:ampl-QED-leading}) we find
\begin{align}\label{eq:emitted_K_QED_pre}
    \langle \mathbb{K}^{\mu} \rangle_{\text{QED}}  &= \alpha^2 \int_0^T\!\ud x^\LCm\ud y^\LCm
    \int_0^{1}\! \ud \sfx\,\frac{\sfx^2}{1-\sfx} \, \int_0^{\infty}\! \ud r \,  r \,  \\
    &\qquad\qquad \times (A_{a}(x^\LCm) A_{a}(y^\LCm)) e^{\frac{i \sfx (m^2+r^2)}{2 n \cdot p(1-\sfx)}  (x^\LCm-y^\LCm)} \frac{2 \left(m^4+r^4\right) }{(n \cdot p)^2 \left(m^2+r^2\right)^2} \tilde{p}^{\mu} \nonumber 
\end{align}
where we have defined
\begin{align}\label{eq:ptilde}
\tilde{p}^{\mu} := \Big(n \cdot p,\, \frac{p_\LCperp^2 + r^2}{2 n \cdot p},\,  p^\LCperp\Big) \,.
\end{align}
Fourier transforming the fields simplifies the $r$-dependence of the integrand, allowing the $\ud r$ integral to be performed exactly.  The final result can be  written, as for the cross-section, as an integral over the frequency of the wave profile:
\begin{align}\label{eq:emitted_K_QED}
    \langle \mathbb{K}^{\mu} \rangle_{\text{QED}} &= \alpha^2 \int_0^{+\infty}\!\ud \omega  \,|E_{a}(\omega)|^2\, \left[ p^{\mu}  F^1_{K,\text{QED}}(\nu(\omega)) + n^{\mu} F^2_{K,\text{QED}}(\nu(\omega)) \right]
\end{align}
where the `form factors' $F^j$ are functions of the dimensionless frequency $\nu$ in~\eqref{eq:frequency-nu} 
\begin{align}\label{eq:formfactors_K_QED}
F^1_{K,\text{QED}}(\nu)&= \frac{(n \cdot p)}{\nu ^4 m^4} \Bigg[\frac{2 \nu  (\nu +1) (2 \nu  (\nu +4)+3)}{(2 \nu +1)^2}-(2 \nu +3) \log \left(2 \nu + 1\right)\Bigg] \,, \\
F^2_{K,\text{QED}}(\nu)&= \frac{(\nu +1) }{\nu ^4 m^2} \Bigg[3 \log \left(2 \nu +1\right)+ \frac{2 \nu  (\nu +1) (2 (\nu -3) \nu -3)}{(2 \nu +1)^2} \Bigg] \,. \nonumber 
\end{align}
To take the classical limit, we assume the wavepacket $\phi(p)$ to be peaked around some classical 4-momentum $p^{\mu}$, and we restore the $\hbar$ dependence of all quantities, expressing the emitted photon momentum $k_\mu$ in terms of wavenumber ${\bar k}_\mu$ as $k_\mu = \hbar {\bar k}_\mu$. From the discussion of length scales, recall \eqref{eq:CED_classical_observables}, we identify the ratio of scales for the classical expansion as\footnote{Strictly, one should perform the frequency integrals in order to properly compare terms in the expansion, but this will just have the effect of replacing $\bar{\omega}$ with some fixed classical frequency characterising the field.}
\be
   \nu = \frac{\lambdabar_C}{\lambdabar} = \frac{\hbar \bar{\omega} n\cdot p}{m^2} \ll 1 \;.
\ee
The classical radiated momentum is then found as
\begin{align}\label{eq:emitted_K_QED_cl}
\langle \mathbb{K}^{\mu} \rangle_{\text{QED}} \Big|_{\hbar = 0}  &= \alpha_{\text{cl}}^2  \frac{8 (n \cdot p)}{3 m^4} p^{\mu}
    \int_0^{\infty}\!\ud \bar{\omega} \,|E_{a}(\bar{\omega})|^2\,,
\end{align}
in which the `classical coupling' is
\be
    \alpha_{\text{cl}} = \frac{e^2}{4 \pi} \;,
\ee
with no factor of $\hbar$. As expected, both the quantum and classical radiated momenta depend only on the absolute value of the electric field strength.

We now consider the radiated momentum flow, which is given by 
\begin{align}
    \langle \mathcalbb{P}^{\mu} \rangle_{\text{QED}} &= \sum_\sigma\int\!\ud \Phi(k)  \hat{\delta}^2(\Omega - \Omega_{\hat{v}})\, k^{\mu} \int\ud\Phi(p') |\mathcal{A}_{3,\sigma}^{(0)}|^2\;.
\end{align}
In order to simplify the geometry, we work in the frame where $p_{\perp} = 0$ so that $r_{\perp} = k_{\perp}/\sfx$ and we can identify the direction $\hat{v}$ on the celestial sphere with spherical coordinates with the natural angles introduced in \eqref{eq:angles}. In particular, we can define a new vector $R^{\mu}$ which is completely determined by the location of the detector $\hat{v} \leftrightarrow (\vartheta,\varphi)$, and in terms of which localized observables are naturally expressed:
\begin{align}\label{eq:rlocalized}
    R^{\perp} &= \sqrt{2} (n \cdot p) \cot \left(\frac{\vartheta }{2}\right) \left( \cos (\varphi), \sin (\varphi)\right) \,, \\
    R^+ &=  \frac{p^{\perp} \cdot  R^{\perp}}{(n \cdot p)} +\frac{|R^{\perp}|^2 -m^2}{2 (n \cdot p)}\,, \quad\quad n \cdot R = 0 \,. \nonumber 
\end{align}
A direction calculation yields
\begin{align}
\label{eq:emitted_P_QED}
    \langle \mathcalbb{P}^{\mu} \rangle_{\text{QED}} &= (2\pi)^2 \frac{\alpha^2}{\pi} \frac{\left(|R^{\perp}|^2+2(n \cdot p)^2\right)^2 }{(n \cdot p) \left(|R^{\perp}|^2+m^2\right)^2}  (p^{\mu}+R^{\mu})  \\
    & \times \int_0^{+\infty}\!\ud \omega \, \frac{1}{\left(|R^{\perp}|^2+(2 \nu(\omega) + 1) m^2\right)^3}  \nonumber \\
    &\times\Big\{(|R^{\perp}|^4+m^4) |E_a(\omega)|^2 -2 m^2 |R^{\perp}|^2 \Big[\cos (2 \varphi) (|E_1(\omega)|^2 -|E_2(\omega)|^2 ) \nonumber \\
    &\qquad\qquad\qquad\qquad\qquad\qquad\qquad\quad + \sin (2 \varphi) (E_1(\omega) E_2^*(\omega)+E_1^*(\omega) E_2(\omega))\Big] \Big\}   \,. \nonumber 
 \end{align}
As for the differential cross-section, the momentum flow carries information about the angular dependence of the emitted radiation. In particular, there is not only a term dependent on the strength $|E_1(\omega)|^2+|E_2(\omega)|^2$ of the electric field but also contributions related to the single components $E_1(\omega),E_2(\omega)$. The classical limit of \eqref{eq:emitted_P_QED} is 
\begin{align}
\label{eq:emitted_P_QED_cl}
    \langle \mathcalbb{P}^{\mu} \rangle_{\text{QED}} &\Big|_{\hbar = 0} = (2\pi)^2 \frac{\alpha_{\text{cl}}^2}{\pi} \frac{\left(|R^{\perp}|^2+2(n \cdot p)^2\right)^2 }{(n \cdot p) \left(|R^{\perp}|^2+m^2\right)^5}  (p^{\mu}+R^{\mu}) \\
    & \times \int_0^{+\infty}\!\ud \bar{\omega} \, \Big\{(|R^{\perp}|^4+m^4) |E_a(\bar{\omega})|^2 -2 m^2 |R^{\perp}|^2 \Big[\cos (2 \varphi) (|E_1(\bar{\omega})|^2 -|E_2(\bar{\omega})|^2 ) \nonumber \\
    &\qquad\qquad\qquad\qquad\qquad\qquad\qquad\quad + \sin (2 \varphi) (E_1(\bar{\omega}) E_2^*(\bar{\omega})+E_1^*(\bar{\omega}) E_2(\bar{\omega}))\Big] \Big\}   \,.  \nonumber
\end{align}

\subsection{The angular momentum flow $\mathcalbb{N}^{\mu \nu}_{\text{QED}}$ and the total angular impulse $J^{\mu \nu}_{\text{QED}}$}

The radiated angular momentum can be computed from the definition
\begin{align}\label{eq:angularmomQED}
    \hspace{-15pt}\langle \mathbb{J}^{\mu \nu} \rangle_{\text{QED}} &= \bra{{\text{in}_\text{QED}}} S^{\dagger} \mathbb{J}^{\mu \nu}_{\text{QED}} S \ket{{\text{in}_\text{QED}}} \\
    &=\int \!\ud \Phi_{\text{out}} \sum_{\sigma} \int \!\ud \Phi(k)  \bra{{\text{in}_\text{QED}}} S^{\dagger} \varepsilon_{\sigma}^{\alpha}(k) a^{\dagger}_{\sigma}(k) \ket{\text{out}} (\mathcal{J}_{\text{QED}})^{\mu \nu}_{\alpha \beta} \bra{\text{out}} \varepsilon_{\sigma}^{*\beta}(k) a_{\sigma}(k) S \ket{{\text{in}_\text{QED}}} \nonumber \\
    &= -i \int \!\ud \Phi(p') \int \!\ud \Phi(k) \eta_{\alpha \beta} \Pi^{\alpha \xi} \bra{{\text{in}_\text{QED}}} S^{\dagger} \ket{\text{out}}_{\xi} \left(k^{[\mu} \frac{\partial}{\partial k_{\nu]}}\right) \Pi^{\beta \zeta} \bra{\text{out}} S \ket{{\text{in}_\text{QED}}}_{\zeta} \nonumber \\
    &\qquad-i\int \!\ud \Phi(p') \int \!\ud \Phi(k) \delta_{\alpha}^{[\mu} \delta_{\beta}^{\nu]}  \Pi^{\alpha \xi} \bra{{\text{in}_\text{QED}}} S^{\dagger} \ket{p', k}_{\xi} \Pi^{\beta \zeta} \bra{p', k} S \ket{{\text{in}_\text{QED}}}_{\zeta} \,, \nonumber 
\end{align}
where we have defined the matrix elements stripped of their polarisation vectors
\begin{align}
\bra{p', k} S \ket{{\text{in}_\text{QED}}} &=: \varepsilon^{\xi}(k) \bra{p', k} S \ket{{\text{in}_\text{QED}}}_{\xi}\,, 
\end{align}
and the light-cone projector
\begin{align}
\Pi^{\alpha \xi} := -\eta^{\alpha \xi} + \frac{k^{\alpha} n^{\xi}+k^{\xi} n^{\alpha}}{k \cdot n} \,.
\end{align}
At leading order, \eqref{eq:angularmomQED} becomes
\begin{align}\label{eq:angularmomQEDlead}
    \langle \mathbb{J}^{\mu \nu}\rangle_{\text{QED}} & = -i \int \!\ud \Phi(p') \int \!\ud \Phi(k) \\
    & \times \Bigg\{\left( \Pi^{\alpha \xi} \frac{\phi^*(p'+k) e^{-i \frac{b \cdot (p'+k)}{\hbar}}}{2 n \cdot (p'+ k)} \mathcal{A}_{3,\xi}^{*(0)}\right) \left(k^{[\mu} \frac{\stackrel{\leftrightarrow}{\partial}}{\partial k_{\nu]}}\right) \left( \Pi_{\alpha}^{\,\, \zeta} \frac{\phi(p'+k) e^{i \frac{b \cdot (p'+k)}{\hbar}}}{2 n \cdot (p'+ k)} \mathcal{A}_{3,\zeta}^{(0)} \right)\nonumber \\
    &\qquad \qquad \qquad \qquad + \delta_{\alpha}^{[\mu} \delta_{\beta}^{\nu]} \frac{|\phi(p'+k)|^2}{4 (n \cdot (p'+ k))^2}  (\Pi^{\alpha \xi} \mathcal{A}_{3,\xi}^{*(0)}) (\Pi^{\beta \zeta} \mathcal{A}_{3,\zeta}^{(0)}) \Bigg\}\,,  \nonumber 
\end{align}
which can now be explicitly evaluated with the amplitude in \eqref{eq:ampl-QED-leading}.  As expected from \eqref{eq:angularmomQEDlead}, the classical $b^{\mu}$ dependence is factored out, and therefore $\langle \mathbb{J}^{\mu \nu}_{\text{QED}} \rangle$ contains a term
\begin{align}\label{eq:b-relation}
    \langle \mathbb{J}^{\mu \nu} \rangle_{\text{QED}} \supset b^{[\mu} \langle \mathbb{K}^{\nu]}\rangle_{\text{QED}} \,.
\end{align}
We therefore obtain
\begin{align}\label{eq:emitted_J_QED}
    \langle \mathbb{J}^{\mu \nu} \rangle_{\text{QED}} &= \alpha^2 \int_0^{\infty}  \!\ud  \omega \, \Big[ |E_a(\omega)|^2 \left(F^1_{J,\text{QED}}(\nu(\omega)) \, b^{[\mu} p^{\nu]} + F^2_{J,\text{QED}}(\nu(\omega)) \, b^{[\mu} n^{\nu]} \right) \\
    &\qquad\qquad\qquad+\frac{F^2_{J,\text{QED}}(\nu(\omega))}{n\cdot p}\,p^{[\mu}n^{\nu]}\Im\!\left[E_a^*(\omega)\partial_\omega E_a(\omega)\right]\nonumber\\
    &\qquad\qquad\qquad-2F^3_{J,\text{QED}}(\nu(\omega))\Im\!\left[E_1^*(\omega)E_2(\omega)\right]\epsilon^{\mu\nu\alpha\beta}n_\alpha p_\beta\Big]\,,\nonumber
\end{align}
in terms of the form factors 
\begin{align}
F^1_{J,\text{QED}}(\nu) &=  F^1_{K,\text{QED}}(\nu)\,, \qquad\qquad\qquad\qquad F^2_{J,\text{QED}}(\nu) =  F^2_{K,\text{QED}}(\nu) \,, \\
F^3_{J,\text{QED}}(\nu) &= \frac{6\nu^2+2\nu-(2\nu+1)^2\log(2\nu+1)}{m^4\nu^4(2\nu+1)}\,. \nonumber 
\end{align}
It is worth noting that, provided  $b^{\mu}$ is only transverse so that $b^- = b^+ = 0$, the components $\langle \mathbb{J}^{a b}_{\text{QED}} \rangle$, $\langle \mathbb{J}^{+ a}_{\text{QED}} \rangle$ and $\langle \mathbb{J}^{- a}_{\text{QED}} \rangle$ can be non-vanishing at this order. The first contribution in \eqref{eq:emitted_J_QED} is related to the standard ``mechanical'' angular momentum due to the position and momentum of the scalar emitting radiation (see \eqref{eq:b-relation}). The last term is more interesting --  it is orthogonal to both the scalar momentum $p^\mu$ and the plane wave direction $n^\mu$-- and it represents a genuine spin-orbit contribution. The classical limit gives 
\begin{align}\label{eq:emitted_J_QED_cl}
    \langle \mathbb{J}^{\mu \nu}\rangle_{\text{QED}}  \Big|_{\hbar = 0} &= \alpha_{\text{cl}}^2 \frac{8 (n \cdot p)}{3 m^4} b^{[\mu} p^{\nu]} \int_0^{+\infty}  \!\ud  \bar{\omega} \,  \,|E_a(\bar{\omega})|^2\\
    & \qquad +\alpha_{\text{cl}}^2 \frac{16 \epsilon^{\mu \nu \alpha \beta} n_{\alpha} p_{\beta}}{3 m^2 (n \cdot p)} \int_0^{+\infty}  \!\ud \bar{\omega} \, \frac{\Im\!\left[E_1^*(\bar{\omega})E_2(\bar{\omega})\right]}{\bar{\omega}} \,.  \nonumber 
\end{align}
We now turn our attention to the angular momentum flow, working in the same frame $p_{\perp} = 0$ as for the momentum flow. We expect that in our geometry the dependence on the wavefunction position will be of the form
\begin{align}\label{eq:b-relation-local}
    \langle \mathcalbb{N}^{\mu \nu} \rangle_{\text{QED}}  &\supset b^{[\mu} \langle \mathcalbb{P}^{\nu]} \rangle_{\text{QED}} \,,
\end{align}
as we showed for the global analogues in \eqref{eq:b-relation}. A straightforward calculation then gives, with $\sfx=2\nu(\omega)m^2/[|R^\perp|^2+(2\nu(\omega)+1)m^2]$ in the integrand,
\begin{align}
\label{eq:emitted_N_QED}
\langle\mathcalbb N^{\mu\nu}\rangle_{\text{QED}}&= -\frac{8\pi \alpha^2\bigl(|R^\perp|^2+2(n\cdot p)^2\bigr)^2|R^\perp|^2}{(n\cdot p)^2(m^2+|R^\perp|^2)^2}\epsilon^{\mu\nu\alpha\beta}\nonumber\\
&\quad\times\int_0^\infty\!\ud \omega\,\frac{\Im\!\left[E_1^*(\omega)E_2(\omega)\right]}{\omega\bigl(|R^\perp|^2+(2\nu(\omega)+1)m^2\bigr)^2}\nonumber\\
&\quad\times\Bigl[(1-\sfx)p_\alpha R_\beta+\frac{(m^2-\sfx|R^\perp|^2)(m^2+|R^\perp|^2)}{2(n\cdot p)|R^\perp|^2}(R_\alpha n_\beta+p_\alpha n_\beta)\Bigr]\nonumber\\
&\quad+\frac{4\pi \alpha^2\bigl(|R^\perp|^2+2(n\cdot p)^2\bigr)^2}{(n\cdot p)^2}p^{[\mu}R^{\nu]}\int_0^\infty\!\ud \omega\,\frac{1}{\bigl(|R^\perp|^2+(2\nu(\omega)+1)m^2\bigr)^3}\nonumber\\
&\quad\times\Im\Bigl[E_a^*(\omega)\Bigl(\partial_{\omega}E_a(\omega)-\frac{4m^2R^aR^b}{(m^2+|R^\perp|^2)^2}\partial_{\omega}E_b(\omega)\Bigr)\Bigr]+b^{[\mu}\langle\mathcalbb P^{\nu]}\rangle_{\text{QED}}\,,
\end{align}
where a more involved tensor structure appears: on top of the mechanical-type of angular momentum contribution due to the radiated momentum flow, there are new non-zero components compared to the global contribution. In particular, the components $\langle \mathcalbb{N}^{a b}_{\text{QED}} \rangle$, $\langle \mathcalbb{N}^{+ a}_{\text{QED}} \rangle$ and $\langle \mathcalbb{N}^{- b}_{\text{QED}} \rangle$ are all non-vanishing at this order. The classical limit of \eqref{eq:emitted_N_QED} yields:
\begin{align}
\label{eq:emitted_N_QED_cl}
\langle\mathcalbb N^{\mu\nu}\rangle_{\text{QED}}\Big|_{\hbar=0}&= -\frac{8\pi \alpha_{\text{cl}}^2\bigl(|R^\perp|^2+2(n\cdot p)^2\bigr)^2|R^\perp|^2}{(n\cdot p)^2(m^2+|R^\perp|^2)^4}\epsilon^{\mu\nu\alpha\beta}\nonumber\\
&\quad\times\int_0^\infty\!\ud \bar\omega\,\frac{\Im\!\left[E_1^*(\bar\omega)E_2(\bar\omega)\right]}{\bar\omega} \Bigl[p_\alpha R_\beta+\frac{m^2(m^2+|R^\perp|^2)}{2(n\cdot p)|R^\perp|^2}(R_\alpha n_\beta+p_\alpha n_\beta)\Bigr]\nonumber\\
&\quad+\frac{4\pi \alpha_{\text{cl}}^2\bigl(|R^\perp|^2+2(n\cdot p)^2\bigr)^2}{(n\cdot p)^2(m^2+|R^\perp|^2)^3}p^{[\mu}R^{\nu]}\int_0^\infty\!\ud \bar\omega\nonumber\\
&\quad\times\Im\Bigl[E_a^*(\bar\omega)\Bigl(\partial_{\bar\omega}E_a(\bar\omega)-\frac{4m^2R^aR^b}{(m^2+|R^\perp|^2)^2}\partial_{\bar\omega}E_b(\bar\omega)\Bigr)\Bigr]+b^{[\mu}\langle\mathcalbb P^{\nu]}\rangle_{\text{QED}}\Big|_{\hbar=0}\,.
\end{align}

\section{Gravitational Compton cross-section and the KLN theorem}
\label{sec:KLNtheorem}

Our gravitational observables are expectation values built from a scattering probability, which is closely related to the cross-section of gravitational Compton scattering. This cross section is however \emph{divergent} at forward scattering, with $ \ud\sigma_{\text{GR}} / \ud \Omega \sim 1/\theta^4$~\cite{Holstein:2006bh,Bjerrum-Bohr:2014lea}. The same issue arises in gauge theories, including \emph{massless} QED\footnote{We remark that mixed photon/graviton Compton processes are also divergent at forward scattering; see~\cite{Bjerrum-Bohr:2014lea} as well as~\cite{Nikishov:1989ds,Nikishov:2010zz,Audagnotto:2022lft}.}. Here the KLN theorem, a hallmark of unitarity in QFT, is widely used to define infrared-safe observables by summing over a complete set of degenerate configurations~\cite{Kinoshita:1962ur} (see also~\cite{Lavelle:2005bt}).

The importance of the collinear divergence in gravitational Compton scattering cannot be overstated: it is a thorny issue for the gravitational S-matrix bootstrap in $d=4$, ultimately related to the $t-$channel pole of the four-point amplitude~\cite{deRham:2022hpx}. In this section, we therefore begin by recovering the standard tree-level Compton amplitude and the corresponding cross-section. We then show that the collinear singularity can only be resolved by summing over forward scattering diagrams, consistent with S-matrix unitarity and generalising the results of~\cite{Frye:2018xjj} from gauge theory to gravity. We will subsequently address the consequences of this result for the definition of global observables in the scattering of particles on gravitational plane-wave backgrounds.

\subsection{Conventional contribution from the tree-level amplitude}

\begin{figure}[t!]
\centering
\includegraphics[width=0.8\textwidth]{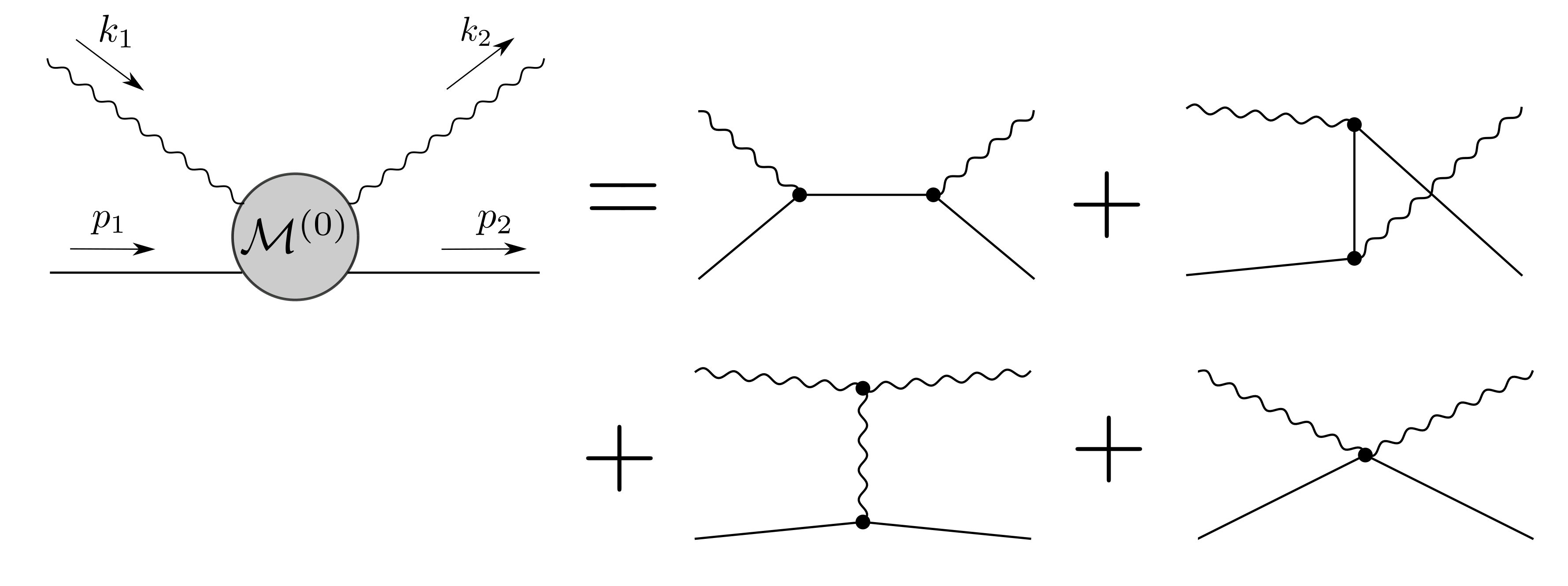}
\caption{The tree-level gravitational Compton amplitude is given by the sum of 4 diagrams, where the $t$-channel contribution (bottom left) is relevant for the collinear divergence.}
\label{fig:tree-Compton}
\end{figure}

\begin{figure}[t!]
\centering
\includegraphics[width=0.60 \textwidth]{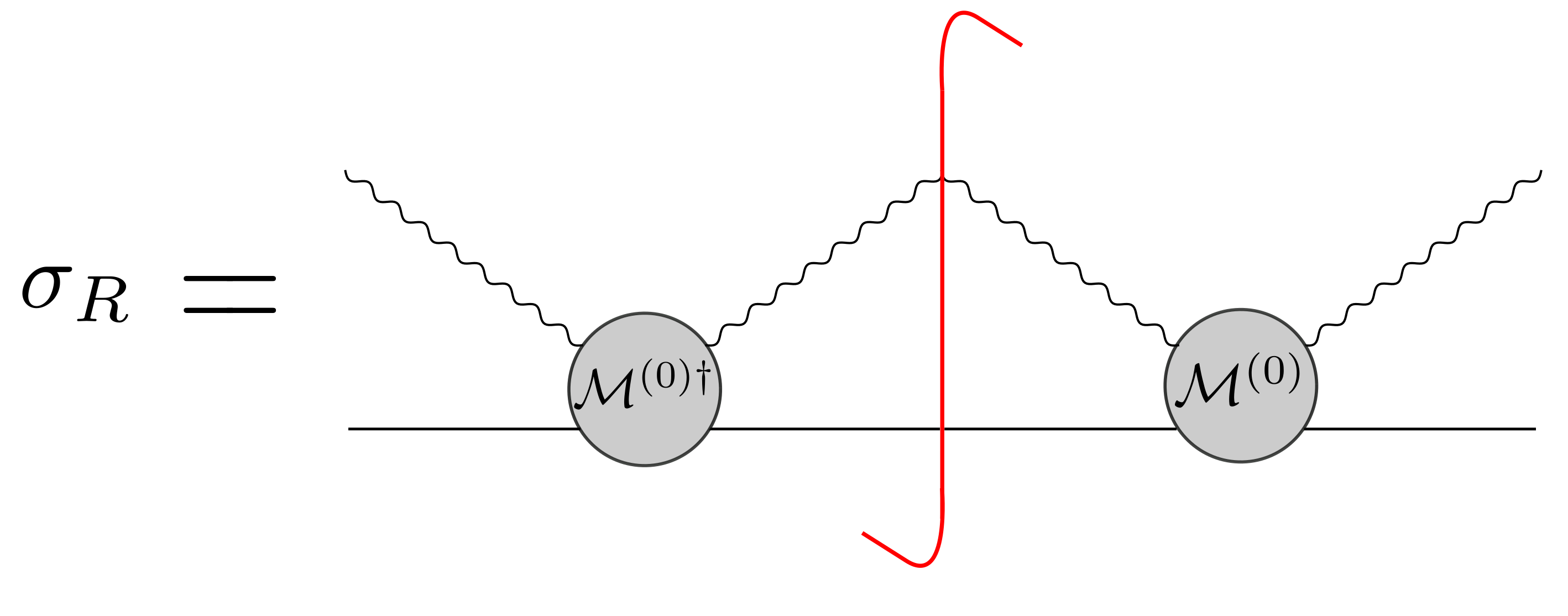}
\caption{The conventional contribution to the cross-section is given by the phase space integration of the tree-level amplitude and its conjugate.}
\label{fig:tree-squared}
\end{figure}

Using a field redefinition and an improved gauge fixing, an efficient framework for both tree and loop-level perturbative quantum gravity calculations {with a minimally coupled scalar} was provided in~\cite{Rafie-Zinedine:2018izq}. We adopt this approach here. The relevant fields are the scalar $\phi$, the graviton $h$ and the ghosts $\chi,\bar{\chi}$. The Lagrangian takes the form
\begin{align}\label{eq:lag1}
\mathcal{L}(h, \phi, \chi, \bar{\chi}) = \mathcal{L}_{\text{grav}}(h) + \mathcal{L}_{\text{matter}}(h, \phi) + \mathcal{L}_{\text{ghost}} (h,\chi, \bar{\chi}) \,,
\end{align}
where, up to the order we are interested in, 
\begin{align}
\mathcal{L}_{\text{grav}}(h) &= \mathcal{L}_{h h h} + \mathcal{L}_{h h h h} + \mathcal{L}_{h h h h h} + \mathcal{O}(\kappa^4)\,, \\
\mathcal{L}_{\text{matter}}(h, \phi) &= \mathcal{L}_{\phi \phi h} + \mathcal{L}_{\phi \phi h h} + \mathcal{L}_{\phi \phi h h h} + \mathcal{O}(\kappa^4)\,, \nonumber \\
\mathcal{L}_{\text{ghost}} (h,\chi,\bar{\chi}) &= \mathcal{L}_{\bar{\chi} \chi h} + \mathcal{L}_{\bar{\chi} \chi h h} + \mathcal{O}(\kappa^3)\,. \nonumber 
\end{align}
We refer the reader to appendix~\ref{sec:appendixA} for the explicit form of the propagator and interaction vertices; we focus here on the calculation itself. The tree-level gravitational Compton amplitude for minimally coupled scalars is given by the sum of $s$, $t$, $u$ channel contributions, and the contact term, as illustrated in Fig.\ref{fig:tree-Compton}. We therefore obtain from the Lagrangian above (see also \cite{Bjerrum-Bohr:2014lea,Holstein:2006bh})
\begin{align}
\mathcal{M}^{(0)}_4 &(p_1,k_1^{\sigma_1};p_2,k_2^{\sigma_2}) = 16 \pi G \frac{(p_1 \cdot k_1) (p_1 \cdot k_2)}{k_1 \cdot k_2}  \\
& \times \left[\frac{(\varepsilon_1^{\sigma_1}(k_1) \cdot p_1) (\varepsilon_2^{*\sigma_2}(k_2) \cdot p_2)}{p_1 \cdot k_1} - \frac{(\varepsilon_1^{\sigma_1}(k_1) \cdot p_2) (\varepsilon_2^{*\sigma_2}(k_2) \cdot p_1)}{p_1 \cdot k_2} - \varepsilon_2^{*\sigma_2}(k_2) \cdot \varepsilon_1^{\sigma_1}(k_1) \right]^2\,, \nonumber
\end{align}
where $p_1$ ($p_2$) is the incoming (outgoing) momentum of the massive particle while $k_1$ ($k_2$) is the incoming (outgoing) graviton momentum. Defining $\overline{|\mathcal{M}^{(0)}_4|^2}$ to be the spin-averaged square of the tree-level amplitude, i.e.
\begin{align}
 \overline{|\mathcal{M}^{(0)}_4|^2} \equiv \frac{1}{2} \sum_{\sigma_1,\sigma_2 = \pm 2} |\mathcal{M}^{(0)}_4(p_1,k_1^{\sigma_1};p_2,k_2^{\sigma_2})|^2\,,
\end{align}
the conventional (`real') contribution to the total scattering cross-section, see Fig.\ref{fig:tree-squared}, is given by the phase space integral
\begin{align}
\label{eq:cross-section-def}
\sigma_R&=  \int \mathrm{d} \Phi(k_2) \int \mathrm{d} \Phi(p_2) \hat{\delta}^d(p_1 + k_1 - p_2 - k_2)   \frac{1}{\mathcal{F}} \overline{|\mathcal{M}^{(0)}_4|^2} \\
&= \int \frac{\mathrm{d}^d k_2}{(2 \pi)^{d-1}} \delta(k_2^2) \Theta(k_2^0) \hat{\delta}(2 p_1 \cdot (k_1 - k_2)-2 k_1 \cdot k_2)  \frac{1}{\mathcal{F}}  \overline{|\mathcal{M}^{(0)}_4|^2} \Big|_{p_2 = p_1 + k_1 - k_2} \,, \nonumber 
\end{align}
in which $\mathcal{F} = 4 (p_1 \cdot k_1)$ is the flux factor and we work here in $d$ dimensions in order to regulate the collinear divergence.
For the actual evaluation of the integrals we choose the rest frame of the massive scalar particle, which means we can parametrize the kinematics as 
\begin{align}
\label{eq:kinematics_rest}
k_1 &= \omega_1 (1,\underbrace{0,\dots,0}_{d-2},1) \,, \qquad\qquad p_1 =m (1,\underbrace{0,\dots,0}_{d-2},0)\,, \\
k_2 &= \omega_2 (1,\sin(\theta),\underbrace{0,\dots,0}_{d-3},\cos(\theta)) \,, \qquad p_2 = (E_2,\underbrace{\vec{p}_2}_{d-1})\,. \nonumber 
\end{align}
Using \eqref{eq:kinematics_rest} in \eqref{eq:cross-section-def} we obtain
\begin{align}
\label{eq:cross-section-rest}
\sigma_R &= \frac{1}{(2 \pi)^{d-2} 4 m \mathcal{F}} \, \int \mathrm{d} \Omega_{d-2} \, \int \mathrm{d} \omega_2\, \omega_2^{d-3} \frac{\delta(\omega_2 - \frac{\omega_1}{1+ \frac{\omega_1}{m}(1- \cos(\theta))})}{1+ \frac{\omega_1}{m} (1- \cos(\theta))} \overline{|\mathcal{M}^{(0)}_4|^2}  \\
&\stackrel{y=\cos(\theta)}{=} \frac{2^{1-d} \pi^{1-\frac{d}{2}}}{m \mathcal{F} \Gamma(\frac{d}{2}-1)} \int_{-1}^{+1} \mathrm{d} y\, \frac{(1-y^2)^{\frac{d-4}{2}}}{\left(1+ \frac{\omega_1}{m} (1- y) \right)^{d-2}} \omega_1^{d-3} \overline{|\mathcal{M}^{(0)}_4|^2} \,. \nonumber 
\end{align}
The differential cross-section, when restricting to $d=4$, is 
\begin{align}
\frac{\mathrm{d} \sigma_R}{\mathrm{d} \Omega} &= G^2 m^2 \left(\frac{\omega_2}{\omega_1}\right)^2
\left( \frac{\sin^8\big(\frac{\theta}{2}\big)+\cos^8\big(\frac{\theta}{2}\big)}{\sin^4\big(\frac{\theta}{2}\big)} \right) \,,
\label{eq:differential-cross_sectionGR}
\end{align}
which agrees exactly with~\cite{Bjerrum-Bohr:2014lea}. The final integration in dim.~reg.~with $d=4-2\epsilon_{\text{IR}}$ gives
\begin{align}\label{eq:cross-section-rest-final}
\sigma_R &= \frac{16 \pi  G^2 m (m+\omega_1)}{\epsilon_{\text{IR}}} \\
&- \frac{4 \pi m G^2}{3 \omega_1^2 (m+2 \omega_1)^3} \Big[-3 m^6-21 m^5 \omega_1+6 (2 \gamma_E -7) m^4 \omega_1^2+6 (9+14 \gamma_E ) m^3 \omega_1^3 \nonumber \\
&\qquad\qquad\qquad\qquad +24 (17+9 \gamma_E ) m^2 \omega_1^4+24 (27+10 \gamma_E ) m \omega_1^5+32 (11+3 \gamma_E ) \omega_1^6\Big] \nonumber \\
&- \frac{4 \pi m G^2}{3 \omega_1^3 } \Big[\left(3 m^4+6 m^3 \omega_1-24 m \omega_1^3+8 \omega_1^4\right) \tanh ^{-1}\left(\frac{\omega_1}{m+\omega_1}\right)  -12 \omega_1^3 (m+\omega_1) \log \left(\frac{\pi }{\omega_1^2}\right)\Big]  \,. \nonumber 
\end{align}
Observe that the collinear divergence comes from the $t-$channel contribution to the gravitational Compton amplitude, and manifests as a $1/\epsilon_{\text{IR}}$ infrared pole in the total cross-section. 

\subsection{One-loop forward amplitude contribution and IR-finiteness}

\begin{figure}[t!]
\centering
\includegraphics[width=0.9 \textwidth]{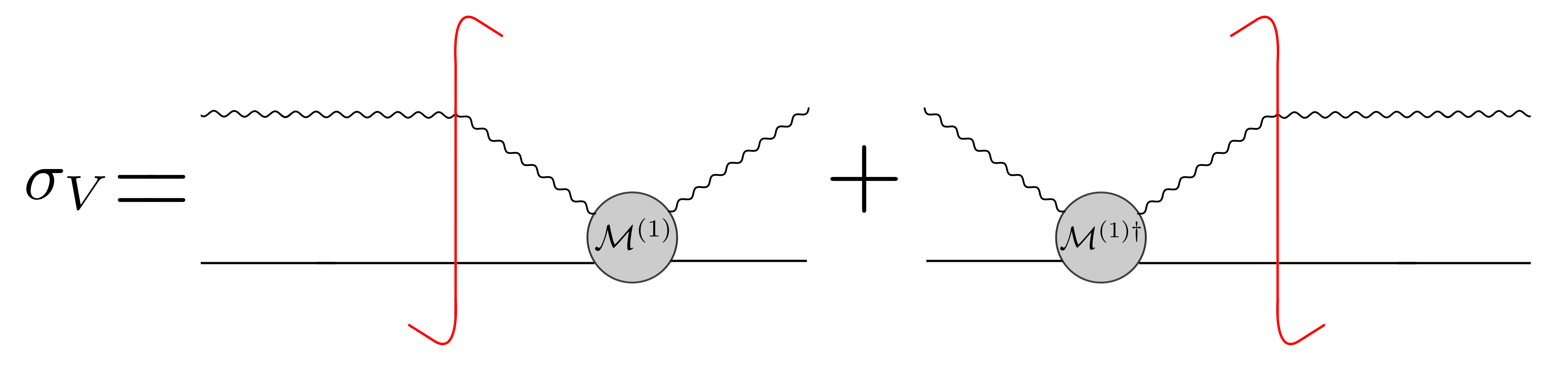}
\caption{The virtual contribution to the cross-section is given by the sum of the phase space integration of the disconnected amplitude and the one-loop forward amplitude.}
\label{fig:one-loop_disconnected}
\end{figure}

According to the KLN theorem, collinear divergences in sufficiently inclusive observables can be removed by summing over degenerate states and processes, the relevant diagrams being required by S-matrix unitarity. In our case, there is a one-loop forward-scattering contribution to the total cross section which is of the same order as the tree-level Compton cross section. We will now show that this generates virtual contributions, localized in the forward region at $t=0$, which cancel the collinear divergence above. While this will solve the problem, it will leave us with an open question from a practical perspective.

\begin{figure}[t!]
\centering
\includegraphics[width=0.98 \textwidth]{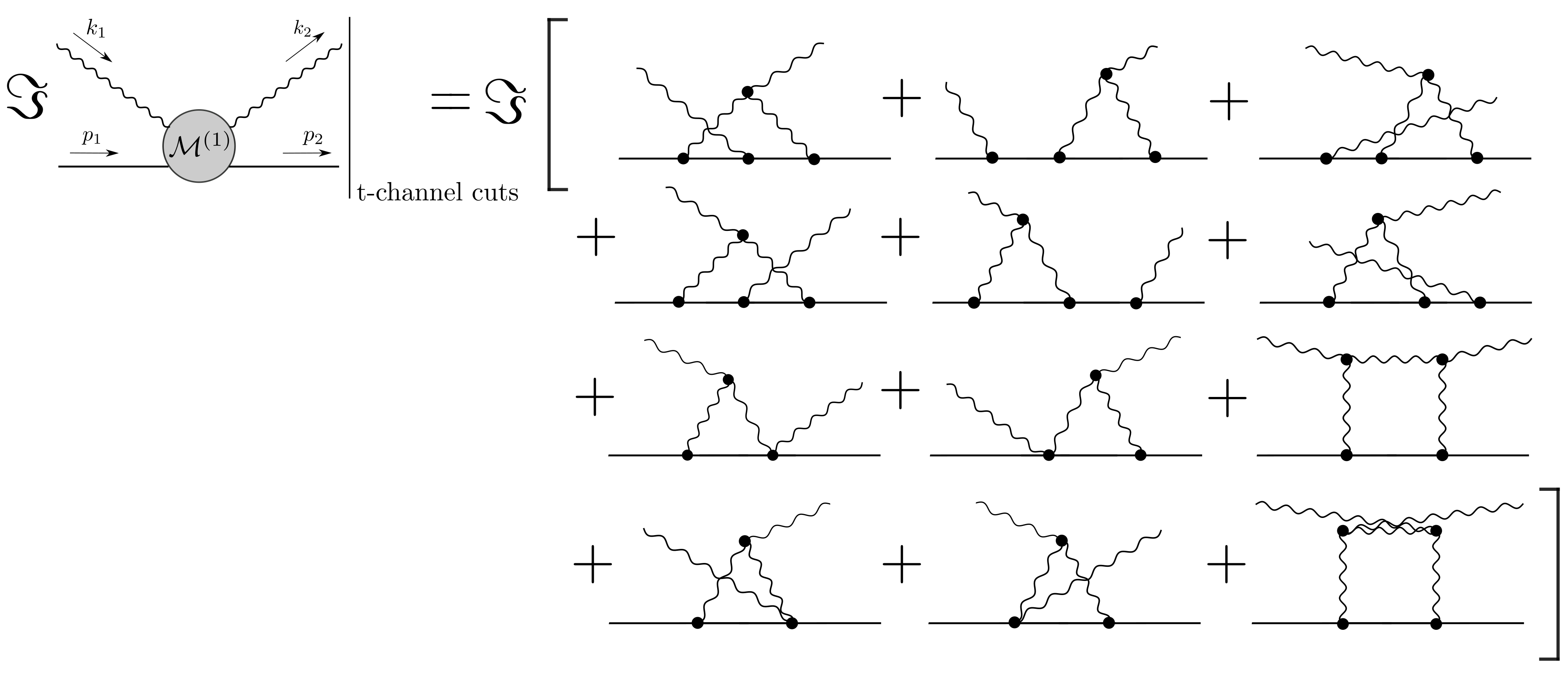}
\caption{The set of diagrams with $t-$channel cuts contributing to the imaginary part of the one-loop Compton amplitude, which are relevant for the collinear divergence.}
\label{fig:tchannel}
\end{figure}

The (virtual) contribution to the cross-section from forward scattering diagrams is shown in Fig.~\ref{fig:one-loop_disconnected} and is given by 
\begin{align}
\sigma_V &= \frac{1}{\mathcal{F}} \int \mathrm{d} \Phi(k_2) \int \mathrm{d} \Phi(p_2) \, 2 \Im \left[ \delta_{\Phi}(p_1 - p_2) \delta_{\Phi}(k_1 - k_2) \overline{\mathcal{M}^{(1)}_4}  \right]
\end{align}
in which $\mathcal{M}^{(1)}_4$ is the one-loop Compton amplitude. Though we may be tempted to write
\begin{align}
\sigma_V &\stackrel{?}{\sim} \frac{1}{\mathcal{F}} 2 \Im \Big( \overline{\mathcal{M}^{(1)}_4} \Big) \Big|_{k_1=k_2,p_1=p_2}\,,
\end{align}
this is ambiguous, because some diagrams include intermediate propagators of the form $\sim {1}/(k_1 - k_2)^2$ which can go on-shell, making the contribution formally divergent\footnote{Evaluating the virtual cross section in this manner would require a detailed analysis for such diagrams which make explicit use of the $i \epsilon$ prescription, as in appendix A of~\cite{Frye:2018xjj}.}. Using FeynArts and FeynCalc with the Lagrangian (\ref{eq:lag1}) we obtain 72 diagrams for the one-loop Compton amplitude. We focus here only on those relevant for the collinear divergence, which are those contributing to the $t$-channel cut and are depicted in Fig.\ref{fig:tchannel}. They are smooth in the forward limit. We compute the sum of those diagrams in dimensional regularization and then extract the infrared-divergent part; we find
\begin{align}
\sigma_V \Big|_{t\text{-channel cuts},\text{IR}} &= \frac{1}{\mathcal{F}} 2 \Im \Big( \overline{\mathcal{M}^{(1)}_4}(p_1, k_1^{\sigma_1};p_1, k_1^{\sigma_1}) \Big)  \Big|_{t\text{-channel cuts},\text{IR}} \\
&=-\frac{16 \pi G^2 m (m + \omega_1)}{\epsilon_{\text{IR}}}\,, \nonumber 
\end{align}
which exactly cancels the infrared-divergent real contribution in~\eqref{eq:cross-section-rest-final},
\begin{align}
\left( \sigma_R + \sigma_V \right) \Big|_{t\text{-channel cuts},\text{IR}} &= 0 \,.
\label{eq:cancellation-IR}
\end{align}
This is the gravitational analogue of the result in Section 6 of~\cite{Frye:2018xjj} for Compton scattering in massless QED, where a similar collinear divergence arises. Unitarity ensures that if all forward scattering terms are included, the final sum of all terms will vanish 
\begin{align}
\sigma_R + \sigma_V = 0 \,.
\label{eq:cancellation}
\end{align}
This can be interpreted as the fact that for hard gravitons (which in our case will compose the gravitational plane-wave background) there is no energy penalty in producing a pair of collinear soft and hard gravitons asymptotically.

The presence of collinear divergences seems to be in contrast with the seminal discussion of Weinberg \cite{Weinberg:1965nx} and recent work on the cancellation of collinear divergences in quantum gravity \cite{Akhoury:2011kq,Naculich:2011ry}, but it is not: it is the Compton amplitude, and not the soft emissions on top of such an amplitude, which is responsible for the divergence\footnote{Indeed, as shown earlier taking the soft limit $\omega_1 \to 0$ implies also $\omega_2 \to 0$: only simultaneous double soft theorems are well-defined in this context~\cite{Klose:2015xoa,Gonzo:2022tjm}.}.

Ultimately, the cancellation in~\eqref{eq:cancellation} will force us to change our physical interpretation of some gravitational backgrounds, such as plane waves. We will discuss this in our calculation of gravitational observables, to which we now turn.

\section{Wave scattering observables in Einstein gravity}\label{sec:GR-obs-section}

In this section we compute the wave scattering observables discussed in section \ref{sec:wavescattering} in Einstein gravity, treating general relativity as an effective field theory valid below the Planck scale. The collinear divergence of the gravitational Compton amplitude implies problems in defining global, infrared-finite observables in gravitational scattering on plane-wave backgrounds. We make this concrete here, by relating observables to the cross section in Section~\ref{sec:KLNtheorem}.

As in electrodynamics it is useful to begin by exploring the relevant length scales. The natural scale for the gravitational interaction is the effective Schwarzschild radius for our problem, i.e. $R_S = G m$. We also have, as in electrodynamics, the Compton wavelength $\lambdabar_C$ and the
classical length scale~$\lambdabar$ defined in \eqref{eq:def_scalesQED}. For quantum observables we will assume\footnote{We note that the condition $G \bar{\omega} (n \cdot p) \ll 1$ follows directly from these inequalities, similar to the condition of validity for the post-Minkowskian expansion.}
\begin{align}
\label{eq:GR_quantum_observables}
R_S \ll \lambdabar_C,\lambdabar \,,
\end{align}
whereas for classical observables we will impose
\begin{align}
\label{eq:GR_classical_observables}
R_S \ll \lambdabar_C \ll \lambdabar \,.
\end{align}

\subsection{Gravitational scattering probability}
Under the assumption that memory effects are negligible\footnote{See~\cite{Cristofoli:2022phh} for a recent discussion of memory in wave-particle scattering.}, the amplitude and probability for graviton emission from a scalar on a gravitational plane wave background have the same overall structure as in (\ref{eq:amp-QED-OK}) and (\ref{eq:QED-P-0}) respectively, for $k,\sigma$ now the momentum and helicity of the emitted graviton; the analogue of $\mathcal{I}_3^\sigma$ in (\ref{eq:amp-mess-qed}) is however more complicated than in QED, and we refer the reader to \cite{Adamo:2017nia,Adamo:2020qru} for details. Here, as in QED, we only need the leading order weak-field expansion of the three-point amplitude to exhibit its relation to the Compton gravitational cross-section; this is
\begin{align}\label{eq:ampl-GR-leading}
\mathcal{M}_{3,\sigma}^{(0)}& = i \kappa^2 \int\!\ud\Phi(p)\, \phi(p)e^{ib\cdot p}
	\hat{\delta}(n \cdot (p'+k - p)) \hat{\delta}^2_{\LCperp}(p'+k - p)  \frac{n \cdot p}{n \cdot k}  (n \cdot (p-k)) \\
&\hspace{-5pt}\times \int_0^T\!\ud x^\LCm\, e^{i \frac{k \cdot p}{n \cdot (p - k)} x^\LCm} \frac{\left[ (\varepsilon_{\sigma}(k) \cdot p) \sfx\, r^a - (k \cdot p) \varepsilon^a_{\sigma}(k) \right]}{k \cdot p} \mathcal{H}_{a b}(x^\LCm) \frac{\left[ (\varepsilon_{\sigma}(k) \cdot p) \sfx\, r^b - (k \cdot p) \varepsilon^b_{\sigma}(k) \right]}{k \cdot p} \,, \nonumber 
\end{align}
in which $\mathcal{H}$ is the integral of the curvature
\begin{align}
\label{eq:GR_waveprofile}
    \mathcal{H}_{a b}(x^\LCm) := \int_0^{x^\LCm} \ud y^\LCm \, H_{a b}(y^\LCm)\,.
\end{align}
We see that \eqref{eq:ampl-GR-leading} has a clear double-copy structure in relation to the QED amplitude \eqref{eq:ampl-QED-leading}: for a summary of double copy relations between gauge and gravitational Compton amplitudes see~\cite{Holstein:2017dwn}. The variables $r_\LCperp$ and $\sfx$ are the same combinations of graviton momentum as used in QED for photon momentum, and the same kinematics holds. As such we present the \emph{differential} scattering probability at leading order in $G^2$,
\be\begin{split}
\label{eq:diff_cross-sectionGR}
	\frac{\ud \mathbb{P}_{\text{GR}}}{\ud \cos\theta} &= 8\int \frac{\ud \omega}{\pi}\frac{I_H(\omega)}{\omega}  \, \frac{\ud\sigma_{\text{GR}}(\nu,\theta)}{\ud\cos\theta} \;,
\end{split}
\ee
in which the gravitational cross section is exactly as in our earlier calculation~\eqref{eq:differential-cross_sectionGR} and we have defined the Fourier transform of the (integrated) wave profile in \eqref{eq:GR_waveprofile}
\begin{align}
    \mathcal{H}_{a b}(\omega) := \int_0^T\!\ud x^\LCm e^{i \omega x^\LCm} \mathcal{H}_{a b}(x^\LCm) \,.
\end{align}
We use the spectral intensity
\begin{align}
I_H(\omega):=|\mathcal H_{11}(\omega)|^2+|\mathcal H_{12}(\omega)|^2=\frac12\mathcal H_{ab}^*(\omega)\mathcal H_{ab}(\omega)\,.
\end{align}
We emphasise that the $1/\theta^4$ divergence of the gravitational Compton cross section at $\theta=0$ is inherited directly by the scattering probability.

Now, recall that $\theta$ is a boosted generalisation of the scattering angle in the frame where the collision is head on. The point $\theta=0$ corresponds to scattering \emph{collinear} with the background, equivalently scattering in the single direction for which the background metric is \emph{not} asymptotically flat; clearly this should be expected to be a subtle limit~\cite{Adamo:2022qci},
and potentially ill-defined. While local (i.e.~differential) observables are well-defined everywhere on the celestial sphere aside from this point, how do we properly define (global) gravitational observables integrated over the sphere? The analysis of section \ref{sec:KLNtheorem} suggests that we should \emph{redefine} the external gravitational wave profile, along the plane-wave direction, in order to absorb the collinear divergence. Equivalently, we should \emph{dress} our incoming state to effectively regulate the gravitational field contribution coming from collinear gravitons interacting at late times; this is similar to what is done for parton distributions in perturbative QCD~\cite{Ellis:1996mzs,White:2022wbr}, and we will borrow these ideas now to address the collinear divergence in gravitational observables. 

\subsection{Coherent state dressing for gravitational Compton observables}
\label{sec:GRobservables}

We propose here a dressing of the initial state, which implements the angular exclusion at this order for a fixed classical trajectory. This is reminiscent of a proposal in the gravitational flat space S-matrix bootstrap program in~\cite{Caron-Huot:2021rmr,Caron-Huot:2021enk}, where a similar regulator was used to tackle the forward limit divergence for practical applications. We define the incoming dressed state as $|\hspace{-1pt}|\text{in}_\text{GR}\rangle\hspace{-2pt}\rangle = \ket{\psi}\otimes|\hspace{-1pt}|\beta\rangle\hspace{-2pt}\rangle$, where
\begin{align}
\label{eq:instate-dressed}
|\hspace{-1pt}|\beta\rangle\hspace{-2pt}\rangle&:= \mathcal{N}_{\beta} \exp\bigg( \sum_\sigma\int \!\ud \Phi(k) \left[\beta^{\sigma}(k) - \beta_{\text{dress}}^{\sigma}(k) \Theta\left(\Lambda_\theta - \Lambda\right)\right] a^{\dagger}_{\sigma}(k)\bigg) \ket{0} \,, 
\end{align}
in which, on top of the gravitational plane-wave background $\beta^{\sigma}(k)$, we include a new perturbative contribution to the waveshape $\beta_{\text{dress}}^{\sigma}(k)$ which receives contributions at, in principle, each order in the coupling,
\begin{align}
  \beta_{\text{dress}}^{\sigma}(k) = G \beta_{\text{dress}}^{(1)\sigma}(k) + G^2 \beta_{\text{dress}}^{(2)\sigma}(k) + \dots \,.
  \label{eq:waveform-collinear}
\end{align}
As indicated by the step function in \eqref{eq:instate-dressed}, $\beta^\sigma_{\text{dress}}$ has support in a narrow cone around the plane-wave direction $\hat{n}$, the size of which is fixed by a universal dimensionless cutoff $\Lambda$ (a detector resolution)
\begin{align}
    \Lambda < \Lambda_\theta =\frac{r}{(n \cdot p)} = \sqrt{2} \cot\left(\frac{\theta}{2}\right)  \,.
\end{align}
For later convenience we also define the dimensionless variable $\mu_{\Lambda}$ as
\begin{align}
    \mu_{\Lambda}:= \frac{n \cdot p}{m} \Lambda  \,.
\end{align}
The next task is to define \eqref{eq:waveform-collinear}, i.e. determine a scheme for the perturbative calculation. We adopt the simplest choice by directly computing the on-shell expectation value of the graviton field at large distances generated by the time evolution of the undressed state $\ket{\text{in}_\text{GR}}$.  The leading order contribution comes from gravitational Compton, i.e.
\begin{align}
    \beta_{\text{dress}}^{(1)\sigma}(k)  &= \int\!\ud\Phi(p') \bra{\text{in}} S^{\dagger} \ket{p'} \bra{p' k^{\sigma}}  S \ket{\text{in}} \\
    &= 32 \pi \int\!\ud\Phi(p)\, \frac12\Theta\!\bigl(n\cdot(p-k)\bigr)\phi^*(p')\phi(p)e^{ib\cdot(p-p')}\frac{n\cdot p}{n\cdot k}  i \int_0^T\!\ud x^\LCm\, e^{i \frac{k \cdot p}{n \cdot (p - k)} x^\LCm}  \nonumber \\
& \quad \times \frac{\left[ (\varepsilon_{\sigma}(k) \cdot p) \sfx\, r^a - (k \cdot p) \varepsilon^a_{\sigma}(k) \right]}{k \cdot p} \mathcal{H}_{a b}(x^\LCm) \frac{\left[ (\varepsilon_{\sigma}(k) \cdot p) \sfx\, r^b - (k \cdot p) \varepsilon^b_{\sigma}(k) \right]}{k \cdot p} \,, \nonumber 
\end{align}
which (up to the external projection) is nothing else than the leading order waveform. \mbox{Here $p'^\mu=p^\mu-k^\mu+\frac{k\cdot p}{n\cdot(p-k)}n^\mu$ is on shell.} For a fixed classical trajectory at this order, this means that we are changing our gravitational field profile at infinity to reabsorb the collinear divergence coming from the graviton self-interaction at late times. For radiative observables at leading order, we use $\Lambda$ as an angular regulator for the phase-space integration.
\subsection{The momentum flow $\mathcalbb{P}^{\mu}_{\text{GR}}$ and the total momentum $K^{\mu}_{\text{GR}}$}
The total radiated momentum outside the cone is given, at leading order, by 
\begin{align}
    \langle\hspace{-2pt}\langle \mathbb{K}^{\mu} \rangle\hspace{-2pt}\rangle_{\text{GR}}  &= \sum_\sigma\int\!\ud \Phi(k)\ud\Phi(p') k^{\mu} |\mathcal{M}_{3,\sigma}^{(0)}|^2 \Theta\left(\Lambda - \Lambda_\theta\right) \; \,,
\end{align}
which, expressing the final state integrals in terms of $\sfx$ and $r$, becomes
\begin{align}\label{eq:emitted_K}
    \langle\hspace{-2pt}\langle \mathbb{K}^{\mu} \rangle\hspace{-2pt}\rangle_{\text{GR}} & = G^2 \int_0^T\!\ud x^\LCm \int_0^T\!\ud y^\LCm
    \int_0^{1}\!\ud\sfx\,(1-\sfx)
    \int_0^{+\infty} \ud r \,  r \, e^{\frac{i \sfx (m^2+r^2)}{2 n \cdot p(1-\sfx)}  (x^\LCm-y^\LCm)}\\
    & \qquad \qquad \times \Theta\left(\Lambda -\Lambda_\theta\right) \frac{32 \pi \left(m^8+r^8\right) (\mathcal{H}_{a b}(x^\LCm) \mathcal{H}^{a b}(y^\LCm))}{\pi \left(m^2+r^2\right)^4} \tilde{p}^{\mu} \,, \nonumber 
\end{align}
where $\tilde{p}^{\mu}$ was defined in \eqref{eq:ptilde}. Inspection confirms that the $r$-integral divergence at large $r$, meaning small scattering angle $\theta$ of the graviton, is indeed regulated by the angular acceptance cut. Physically, this means that the observer is only looking at the portion of the sky which is complementary to the cone of size $\Lambda$ around the plane-wave direction $\hat{n}$.  Performing the integrals, the final result can be written
\begin{align}\label{eq:emitted_K_GR}
    \langle\hspace{-2pt}\langle \mathbb{K}^{\mu} \rangle\hspace{-2pt}\rangle_{\text{GR}} &= G^2 \int_0^{\infty}\!\ud \omega  I_H(\omega) \Big[p^{\mu}  F^1_{K,\text{GR}}(\nu(\omega)) + n^{\mu} F^2_{K,\text{GR}}(\nu(\omega))\Big]
\end{align}
in terms of the form factors 
\begin{align}\label{eq:formfactors_K_GR}
    F^1_{K,\text{GR}} (\nu)&:= \frac{8 (n \cdot p)}{\nu ^4} \Big[\left(8 \nu ^4+4 \nu +3\right) \log \left(\frac{\mu_{\Lambda}^2+2 \nu +1}{2 \nu +1}\right)-(4 \nu +3) \log \left(\mu_{\Lambda}^2+1\right) \nonumber  \\
    & \qquad +\frac{\left(\mu_{\Lambda}^2+1\right) \left(\mu_{\Lambda}^{10}+\mu_{\Lambda}^8 (4 \nu -3)-4 \mu_{\Lambda}^6 (2 \nu +1)+5 \mu_{\Lambda}^2+12 \nu +5\right)}{4 \left(\mu_{\Lambda}^2+2 \nu +1\right)^2} \nonumber \\
 & \qquad-\frac{\mu_{\Lambda}^8}{4}+\mu_{\Lambda}^6+\mu_{\Lambda}^4 \nu ^2+2 \mu_{\Lambda}^2 \nu ^2 (1-2 \nu )-\frac{2 \mu_{\Lambda}^2 \left(\mu_{\Lambda}^4-2\right) \nu }{\mu_{\Lambda}^2+1}-\frac{12 \nu +5}{4 (2 \nu +1)^2}\Big]\,, \nonumber \\
    F^2_{K,\text{GR}} (\nu)&:= \frac{8 m^2}{\nu ^4} \Big[-2 \left(24 (\nu +1) \nu ^4+5 \nu +3\right) \arccoth\left(\frac{\mu_{\Lambda}^2+4 \nu +2}{\mu_{\Lambda}^2}\right)\nonumber \\
 & \qquad +(5 \nu +3) \log \left(\mu_{\Lambda}^2+1\right)+\frac{7 \nu +3}{4 (2 \nu +1)^2} -\frac{\mu_{\Lambda}^2 \left(\mu_{\Lambda}^8+5 \mu_{\Lambda}^6-4 \mu_{\Lambda}^4+16\right) \nu }{4 \left(\mu_{\Lambda}^2+1\right)} \nonumber \\
    & \qquad +\frac{\left(\mu_{\Lambda}^2+1\right) \left(\mu_{\Lambda}^2 \left(\mu_{\Lambda}^8 (\nu -3)-\mu_{\Lambda}^6 (7 \nu +1)+\mu_{\Lambda}^4 (4 \nu +2)+2 \mu_{\Lambda}^2+5 \nu -1\right)-7 \nu -3\right)}{4 \left(\mu_{\Lambda}^2+2 \nu +1\right)^2}\nonumber \\
 & \qquad +\frac{1}{4} \mu_{\Lambda}^2 \left(3 \mu_{\Lambda}^6+4 \left(\mu_{\Lambda}^4-3\right) \nu ^2-2 \mu_{\Lambda}^4-12 \left(\mu_{\Lambda}^2-2\right) \nu ^3+48 \nu ^4-2\right)\Big]\,. 
\end{align}
Taking the classical limit proceeds as usual, by restoring the dependence on $\hbar$ with the effective replacement $\nu \to \hbar \bar{\nu}$ and $G \to G /\hbar$ \cite{Kosower:2018adc} and we impose the hierarchy of scales~ \eqref{eq:GR_classical_observables}. This yields the classical radiated momentum 
\begin{align}
\label{eq:emitted_K_GR_cl}
    \langle\hspace{-2pt}\langle \mathbb{K}^{\mu} \rangle\hspace{-2pt}\rangle_{\text{GR}} &\Big|_{\hbar = 0} = G^2  \frac{32 (n \cdot p)}{3 \left(\mu_{\Lambda}^2+1\right)^4} \Big\{p^{\mu} \left(6 \left(\mu_{\Lambda}^2+1\right)^4 \log \left(\mu_{\Lambda}^2+1\right)-\mu_{\Lambda}^4 \left(11 \mu_{\Lambda}^4+20 \mu_{\Lambda}^2+12\right)\right)  \nonumber \\
    &\qquad+ n^{\mu} \frac{m^2}{(n \cdot p)} \left(15 \mu_{\Lambda}^2-18 \left(\mu_{\Lambda}^2+1\right)^4 \log \left(\mu_{\Lambda}^2+1\right)+\left(\mu_{\Lambda}^2+10\right) \left(3 \mu_{\Lambda}^4+7 \mu_{\Lambda}^2+6\right) \mu_{\Lambda}^4\right) \Big\} \nonumber \\
    &\qquad \times  \int_0^{\infty}\!\ud \bar{\omega}   I_H(\bar{\omega})
\end{align}
Compared to the analogous result in electrodynamics \eqref{eq:emitted_K_QED_cl}, we see here that a component along the plane-wave direction $n^{\mu}$ survives the classical limit: this is related to the presence of classical tail effects, which are purely of gravitational origin.

The momentum flow $\langle\hspace{-2pt}\langle \mathcalbb{P}^{\mu} \rangle\hspace{-2pt}\rangle_{\text{GR}}$ can be computed similarly, but in this case we choose our detector (i.e.~the angle $\Omega_{\hat{v}}$) to be in the complement of the cone around the plane wave direction. Therefore, local observables will not be sensitive to the dressing of the incoming state. We then have for the momentum flow
\begin{align}
    \langle\hspace{-2pt}\langle \mathcalbb{P}^{\mu} \rangle\hspace{-2pt}\rangle_{\text{GR}} &= \langle \mathcalbb{P}^{\mu} \rangle_{\text{GR}} = \sum_\sigma\int\!\ud \Phi(k)\ud\Phi(p') k^{\mu} \hat{\delta}^2(\Omega - \Omega_{\hat{v}}) |\mathcal{M}_{3,\sigma}^{(0)}|^2\;,
\end{align}
and, working in the frame where $p_{\perp} = 0$ as in QED, we find 
\begin{align}\label{eq:emitted_P_GR}
    \langle\hspace{-2pt}\langle \mathcalbb{P}^{\mu} \rangle\hspace{-2pt}\rangle_{\text{GR}} &= (2\pi)^2 \frac{G^2}{\pi} \frac{8 \left(|R^{\perp}|^2+2(n \cdot p)^2\right)^2 }{(n \cdot p) \left(m^2+|R^{\perp}|^2\right)^2 } (p^{\mu}+R^{\mu}) \\
    & \times  \int_0^{+\infty}\!\ud \omega \, \frac{1}{\left(|R^{\perp}|^2+(2 \nu(\omega) + 1) m^2\right)^3} \nonumber \\
    & \times \Big\{(m^8+|R^{\perp}|^8) I_H(\omega) + 2 |R^{\perp}|^4 m^4 \Big[ \sin (4 \varphi)  (\mathcal{H}_{11}(\omega) \mathcal{H}_{12}^*(\omega)+\mathcal{H}_{11}^*(\omega) \mathcal{H}_{12}(\omega)) \nonumber \\
    & \qquad \qquad \qquad \qquad \qquad \qquad  \qquad \qquad \qquad \quad+\cos (4 \varphi) (|\mathcal{H}_{11}(\omega)|^2 - |\mathcal{H}_{12}(\omega)|^2) \Big] \Big\} \nonumber  \,,
\end{align}
where $R^{\mu}=  r^{\mu}(\vartheta,\varphi)$ is defined as in \eqref{eq:rlocalized}. We notice that there are remarkable similarities between \eqref{eq:emitted_K_QED} and \eqref{eq:emitted_K_GR} and between \eqref{eq:emitted_P_QED} and \eqref{eq:emitted_P_GR}, which are a consequence of the double copy of the 3-pt amplitude discussed earlier. The classical limit of \eqref{eq:emitted_P_GR} gives
\begin{align}\label{eq:emitted_P_GR_cl}
    \hspace{-6pt}\langle\hspace{-2pt}\langle &\mathcalbb{P}^{\mu} \rangle\hspace{-2pt}\rangle_{\text{GR}} \Big|_{\hbar = 0} = (2\pi)^2 \frac{G^2}{\pi}  \frac{8 \left(|R^{\perp}|^2+2(n \cdot p)^2\right)^2 }{(n \cdot p) \left(m^2+|R^{\perp}|^2\right)^5 } (p^{\mu}+R^{\mu})  \\
    & \times  \int_0^{+\infty}\!\ud \bar{\omega} \, \Big\{(m^8+|R^{\perp}|^8) I_H(\bar{\omega}) + 2 |R^{\perp}|^4 m^4 \Big[ \sin (4 \varphi)  (\mathcal{H}_{11}(\bar{\omega}) \mathcal{H}_{12}^*(\bar{\omega})+\mathcal{H}_{11}^*(\bar{\omega}) \mathcal{H}_{12}(\bar{\omega})) \nonumber \\
    & \qquad \qquad \qquad \qquad \qquad \qquad  \qquad \qquad \qquad \qquad \qquad \quad+\cos (4 \varphi) (|\mathcal{H}_{11}(\bar{\omega})|^2 - |\mathcal{H}_{12}(\bar{\omega})|^2) \Big] \Big\} \,. \nonumber 
\end{align}

\subsection{The angular momentum flow $\mathcalbb{N}^{\mu \nu}_{\text{GR}}$ and the total angular impulse $J^{\mu \nu}_{\text{GR}}$}

The radiated angular momentum for the gravitational radiation is given by
\begin{align}\label{eq:angularmomGR}
    \langle\hspace{-2pt}\langle \mathbb{J}^{\mu \nu} \rangle\hspace{-2pt}\rangle_{\text{GR}} &= \langle\hspace{-2pt}\langle \text{in}_\text{GR} |\hspace{-1pt}| S^{\dagger} \mathbb{J}^{\mu \nu}_{\text{GR}} S |\hspace{-1pt}| \text{in}_\text{GR} \rangle\hspace{-2pt}\rangle  \\
    &=\int \!\ud \Phi_{\text{out}} \sum_{\sigma} \int \!\ud \Phi(k)  \Theta\left(\Lambda -\Lambda_\theta\right)  \nonumber \\
    &\qquad \times \langle\hspace{-2pt}\langle \text{in}_\text{GR} |\hspace{-1pt}| S^{\dagger} \varepsilon_{\sigma}^{\alpha \alpha'}(k) a^{\dagger}_{\sigma}(k) \ket{\text{out}}  (\mathcal{J}_{\text{GR}})^{\mu \nu}_{\alpha \alpha' \beta \beta'} \bra{\text{out}} \varepsilon_{\sigma}^{*\beta \beta'}(k) a_{\sigma}(k) S |\hspace{-1pt}| \text{in}_\text{GR} \rangle\hspace{-2pt}\rangle \,. \nonumber
\end{align}
At leading order, from \eqref{eq:angularmomGR} we get
\begin{align}
    \hspace{-12pt}\langle\hspace{-2pt}&\langle \mathbb{J}^{\mu \nu} \rangle\hspace{-2pt}\rangle_{\text{GR}} = +i \int \!\ud \Phi(p') \int \!\ud \Phi(k) \Theta\left(\Lambda -\Lambda_\theta\right)  \\
    &\times \Big\{ \Bigg( \Pi^{\alpha \alpha' \xi \xi'} \frac{\phi^*(p'+k) e^{-i \frac{b \cdot (p'+k)}{\hbar}}}{2 n \cdot (p'+ k)} \mathcal{M}_{3,\xi \xi'}^{*(0)}\Bigg) \Bigg(k^{[\mu} \frac{\stackrel{\leftrightarrow}{\partial}}{\partial k_{\nu]}}\Bigg) \Bigg( \Pi_{\alpha \alpha'}^{\quad \zeta \zeta'} \frac{\phi(p'+k) e^{i \frac{b \cdot (p'+k)}{\hbar}}}{2 n \cdot (p'+ k)} \mathcal{M}_{3, \zeta \zeta'}^{(0)} \Bigg) \nonumber \\
    &\qquad\qquad\qquad+2 \eta_{\alpha' \beta'} \delta_{\alpha}^{[\mu} \delta_{\beta}^{\nu]} \frac{|\phi(p'+k)|^2}{4 (n \cdot (p'+ k))^2}  (\Pi^{\alpha \alpha' \xi \xi'} \mathcal{M}_{3,\xi \xi'}^{*(0)}) (\Pi^{\beta \beta' \zeta \zeta'} \mathcal{M}_{3,\zeta \zeta'}^{(0)}) \Big\} \,,\nonumber 
\end{align}
\vspace{-3pt}
where we have defined the matrix elements stripped from their polarisation vectors
\begin{align}
\bra{p', k} S |\hspace{-1pt}| \text{in}_\text{GR} \rangle\hspace{-2pt}\rangle &=: \varepsilon^{\zeta \zeta'}(k) \bra{p', k} S |\hspace{-1pt}| \text{in}_\text{GR} \rangle\hspace{-2pt}\rangle_{\zeta \zeta'}\,,
\end{align}
and the gravitational light-cone projector \cite{Matsuki:1978rt}
\begin{align}
\Pi^{\alpha \alpha' \xi \xi'} := \frac{1}{2}\left(\Pi^{\alpha \xi} \Pi^{\alpha' \xi'} + \Pi^{\alpha \xi'} \Pi^{\alpha' \xi} - \Pi^{\alpha \alpha'} \Pi^{\xi \xi'} \right) \,.
\end{align}
The final radiated angular momentum has a compact expression:
\begin{align}
    &\langle\hspace{-2pt}\langle \mathbb{J}^{\mu \nu} \rangle\hspace{-2pt}\rangle_{\text{GR}} = G^2 \int_0^{+\infty}  \!\ud  \omega \, \Big[I_H(\omega) \left(F^1_{J,\text{GR}}(\nu(\omega)) b^{[\mu} p^{\nu]}  + F^2_{J,\text{GR}}(\nu(\omega)) b^{[\mu} n^{\nu]} \right) \\
    &\qquad+\frac{F^2_{J,\text{GR}}(\nu(\omega))}{2(n\cdot p)}\,p^{[\mu}n^{\nu]}\Im\!\left[\mathcal H_{ab}^*(\omega)\partial_\omega\mathcal H_{ab}(\omega)\right] + \Im\!\left[\mathcal H_{11}^*(\omega)\mathcal H_{12}(\omega)\right] 2 F^3_{J,\text{GR}}(\nu(\omega)) \epsilon^{\mu \nu \alpha \beta} n_{\alpha} p_{\beta}\Big] \,, \nonumber 
\end{align}
where the form factors are given by
\begin{align}
&F^1_{J,\text{GR}}(\nu) =  F^1_{K,\text{GR}}(\nu) \,,\qquad\qquad\qquad F^2_{J,\text{GR}}(\nu) =  F^2_{K,\text{GR}}(\nu) \,,  \\
&F^3_{J,\text{GR}}(\nu) = \frac{32}{\nu^4}\Bigg[\frac{4\nu+1}{2}\log(\mu_\Lambda^2+1)-(32\nu^4+16\nu^3+4\nu+1)\arccoth\!\bigg(\frac{\mu_\Lambda^2+4\nu+2}{\mu_\Lambda^2}\bigg)\nonumber\\
    &\qquad\qquad+2\mu_\Lambda^2\nu^3-\frac{\mu_\Lambda^2\nu^2(2\nu+1)^2(\mu_\Lambda^2+4\nu+2)}{(\mu_\Lambda^2+2\nu+1)^2}+\frac{\mu_\Lambda^2\nu(40\nu^4+48\nu^3+12\nu^2-4\nu-1)}{(2\nu+1)(\mu_\Lambda^2+2\nu+1)}\Bigg]\,. \nonumber
\end{align}
The classical radiated angular momentum is
\begin{align}
    \langle\hspace{-2pt}\langle \mathbb{J}^{\mu \nu} \rangle\hspace{-2pt}\rangle_{\text{GR}} \Big|_{\hbar = 0} &=  G^2 \frac{32 (n \cdot p)}{3} \Bigg[ b^{[\mu} p^{\nu]} \left(6 \log \left(\mu_{\Lambda}^2+1\right)-\frac{\mu_{\Lambda}^4 \left(11 \mu_{\Lambda}^4+20 \mu_{\Lambda}^2+12\right)}{ \left(\mu_{\Lambda}^2+1\right)^4}\right) \nonumber  \\ 
    &\quad \quad+ \frac{m^2}{(n \cdot p)}  b^{[\mu} n^{\nu]} \Big(\frac{15 \mu_{\Lambda}^2 + \left(\mu_{\Lambda}^2+10\right) \left(3 \mu_{\Lambda}^4+7 \mu_{\Lambda}^2+6\right) \mu_{\Lambda}^4}{ \left(\mu_{\Lambda}^2+1\right)^4}-18 \log \left(\mu_{\Lambda}^2+1\right) \Big) \Bigg] \nonumber \\
    & \qquad \times \int_0^{+\infty}  \!\ud \bar{\omega} \, I_H(\bar{\omega}) \nonumber \\
    &+\frac{16G^2m^2}{3(n\cdot p)}\,p^{[\mu}n^{\nu]}\Bigg[\frac{15\mu_\Lambda^2+(\mu_\Lambda^2+10)(3\mu_\Lambda^4+7\mu_\Lambda^2+6)\mu_\Lambda^4}{(\mu_\Lambda^2+1)^4}-18\log(\mu_\Lambda^2+1)\Bigg]\nonumber\\
    &\qquad\times\int_0^\infty\!\ud\bar\omega\,\Im\!\left[\mathcal H_{ab}^*(\bar\omega)\partial_{\bar\omega}\mathcal H_{ab}(\bar\omega)\right]\nonumber\\
    & + G^2 \frac{128 m^2 }{3 (n \cdot p)}  \epsilon^{\mu \nu \alpha \beta} n_{\alpha} p_{\beta}  \Bigg[\frac{\mu_{\Lambda}^2 \left(3 \mu_{\Lambda}^6+23 \mu_{\Lambda}^4+33 \mu_{\Lambda}^2+15\right)}{\left(\mu_{\Lambda}^2+1\right)^3}-12 \log \left(\mu_{\Lambda}^2+1\right)\Bigg] \nonumber \\
    & \qquad \times  \int_0^{+\infty}  \!\ud \bar{\omega} \, \frac{\Im\!\left[\mathcal H_{11}^*(\bar{\omega})\mathcal H_{12}(\bar{\omega})\right]}{ \bar{\omega}} \,.
\end{align}
For the angular momentum flow, we can follow a procedure similar to the electromagnetic case and take advantage of the fact that \eqref{eq:b-relation-local} holds also in the gravitational case.  We then obtain, in the $p_{\perp} = 0$ frame, the following result 
 \begin{align}
\label{eq:emitted_N_GR}
\langle\hspace{-2pt}\langle\mathcalbb N^{\mu\nu}\rangle\hspace{-2pt}\rangle_{\text{GR}}&= -\frac{128\pi G^2\bigl(|R^\perp|^2+2(n\cdot p)^2\bigr)^2|R^\perp|^6}{(n\cdot p)m^2(m^2+|R^\perp|^2)^2}\epsilon^{\mu\nu\alpha\beta}\nonumber\\
&\quad\times\int_0^\infty\!\ud \omega\,\frac{\Im\!\left[\mathcal H_{11}^*(\omega)\mathcal H_{12}(\omega)\right]}{\nu(\omega)\bigl(|R^\perp|^2+(2\nu(\omega)+1)m^2\bigr)^2}\nonumber\\
&\quad\times\Bigl[(1-\sfx)p_\alpha R_\beta+\frac{m^8+m^2|R^\perp|^6-\sfx(m^2+|R^\perp|^2)|R^\perp|^6}{2(n\cdot p)|R^\perp|^6}(R_\alpha n_\beta+p_\alpha n_\beta)\Bigr]\nonumber\\
&\quad+\frac{16\pi G^2\bigl(|R^\perp|^2+2(n\cdot p)^2\bigr)^2}{(n\cdot p)^2}p^{[\mu}R^{\nu]}\int_0^\infty\!\ud \omega\,\frac{1}{\bigl(|R^\perp|^2+(2\nu(\omega)+1)m^2\bigr)^3}\nonumber\\
&\quad\times\Im\Bigl[(m^2-|R^\perp|^2)^2\mathcal H_{ab}^*(\omega)\partial_{\omega}\mathcal H_{ab}(\omega)\nonumber\\
&\qquad\qquad+\frac{8m^4R^aR^bR^cR^d}{(m^2+|R^\perp|^2)^2}\mathcal H_{ab}^*(\omega)\partial_{\omega}\mathcal H_{cd}(\omega)\Bigr]+b^{[\mu}\langle\hspace{-2pt}\langle\mathcalbb P^{\nu]}\rangle\hspace{-2pt}\rangle_{\text{GR}}\,,
\end{align}
where we remind that $R^{\mu}=  r^{\mu}(\vartheta,\varphi)$ is defined as in \eqref{eq:rlocalized}. As noticed earlier for the momentum flow, we notice that the double copy of the amplitude manifest itself in a remarkable similarity between the electromagnetic \eqref{eq:emitted_N_QED} and gravitational \eqref{eq:emitted_N_GR} result. The classical limit of \eqref{eq:emitted_N_GR} yields:
\begin{align}
\label{eq:emitted_N_GR_cl}
\langle\hspace{-2pt}\langle\mathcalbb N^{\mu\nu}\rangle\hspace{-2pt}\rangle_{\text{GR}}\Big|_{\hbar=0}&= -\frac{128\pi G^2\bigl(|R^\perp|^2+2(n\cdot p)^2\bigr)^2|R^\perp|^6}{(n\cdot p)^2(m^2+|R^\perp|^2)^4}\epsilon^{\mu\nu\alpha\beta}\nonumber\\
&\quad\times\int_0^\infty\!\ud \bar\omega\,\frac{\Im\!\left[\mathcal H_{11}^*(\bar\omega)\mathcal H_{12}(\bar\omega)\right]}{\bar\omega} \Bigl[p_\alpha R_\beta+\frac{m^8+m^2|R^\perp|^6}{2(n\cdot p)|R^\perp|^6}(R_\alpha n_\beta+p_\alpha n_\beta)\Bigr]\nonumber\\
&\quad+\frac{16\pi G^2\bigl(|R^\perp|^2+2(n\cdot p)^2\bigr)^2}{(n\cdot p)^2(m^2+|R^\perp|^2)^3}p^{[\mu}R^{\nu]}\int_0^\infty\!\ud \bar\omega\nonumber\\
&\quad\times\Im\Bigl[(m^2-|R^\perp|^2)^2\mathcal H_{ab}^*(\bar\omega)\partial_{\bar\omega}\mathcal H_{ab}(\bar\omega)\nonumber\\
&\qquad\qquad+\frac{8m^4R^aR^bR^cR^d}{(m^2+|R^\perp|^2)^2}\mathcal H_{ab}^*(\bar\omega)\partial_{\bar\omega}\mathcal H_{cd}(\bar\omega)\Bigr]+b^{[\mu}\langle\hspace{-2pt}\langle\mathcalbb P^{\nu]}\rangle\hspace{-2pt}\rangle_{\text{GR}}\Big|_{\hbar=0}\,.
\end{align}
\section{The high-energy limit of wave scattering observables}
\label{sec:highenergy}

In this section we study the behaviour of our classical and quantum radiative observables in the high-energy limit. There are several motivations for this beyond intrinsic interest. For the classical two-body problem, the power emitted has power-law mass singularities for massive point charges in electrodynamics at $\alpha_{\text{cl}}^3$ \cite{Saketh:2021sri,Bern:2021xze} and logarithmic mass divergences for point masses in general relativity at $G^3$ \cite{Herrmann:2021tct,Mougiakakos:2021ckm,Jakobsen:2021smu,DiVecchia:2021bdo,DiVecchia:2022nna,Dlapa:2022lmu,Dlapa:2023hsl,Bini:2022enm,Damour:2022ybd}. Similar mass singularities appear in other classical observables, such as the angular momentum and the scattering angle, raising questions about the regime of validity of the classical expansion~\cite{Kovacs:1977uw,Kovacs:1978eu,DEath:1976bbo,Gruzinov:2014moa,DiVecchia:2022nna}.

In our framework of a point particle emitting radiation in a plane-wave background we will see similar mass singularities arise much earlier in the perturbative expansion. This motivates us to reconsider the high-energy limit from the perspective of the full quantum theory, to see whether this can offer a solution. Geometrically, the high-energy limit (denoted H.E.) is characterised by either\footnote{We can phrase the discussion entirely in terms of dimensionless variables by defining the rapidity $\gamma=(n \cdot p)/m$, so that $\gamma >\!\!> 1$ corresponds to the high-energy limit. Given that we want to isolate the mass singularities, though, we prefer to keep an explicit dependence on $m$.} $n\cdot p\to\infty$, in which the wave-particle collision is almost head-on, or $n\cdot p\to 0$, in which the collision is almost collinear. The fact that there are two options is due to $n_\mu$ being a null vector. We focus on the former case, which is more natural and intuitive; the latter case gives similar results.

\subsection{Classical theory: power-law and logarithmic mass divergences}
We begin in the classical theory, taking the high-energy limit of the radiated lightfront energy $\langle n\cdot \mathbb{K}\rangle$ in QED  (\ref{eq:emitted_K_QED_cl}) and in gravity (\ref{eq:emitted_K_GR_cl}). This is the simplest of our global observables, and its behaviour typifies that of \emph{several} of our observables. In classical electrodynamics we find
\be
    \langle n \cdot \mathbb{K} \rangle_{\text{QED}} \Big|_{\hbar = 0}   \stackrel{\text{H.E.}}{=}  \frac{8 \alpha_{\text{cl}}^2 (n \cdot p)^2}{3 m^4} 
    \int_0^{\infty}\!\ud \bar{\omega} \,|E_{a}(\bar{\omega})|^2\,.
\ee
We notice that the field dependence has factored out and the frequency integral is clearly finite for smooth wave shapes, e.g.~Gaussian type profiles. We conclude that the classical radiated energy grows quadratically in the high energy limit $(n \cdot p)\to\infty$, and exhibits a power-law singularity in the massless limit. In general relativity, the situation is similar. If we restrict our attention to the high-energy behaviour, we obtain
\begin{align}
    \langle n \cdot \mathbb{K}\rangle_{\text{GR}}\bigg|_{\hbar =0}
    \stackrel{\text{H.E.}}{\sim} 64 G^2 (n \cdot p)^2  \bigg( \log\frac{(n\cdot p)^2 \Lambda^2}{m^2}-\frac{11}{6}\bigg) \int_0^{\infty}\!\ud \bar{\omega}\, I_H(\bar\omega)\,,
\end{align}
which grows as $(n\cdot p)^2\log n\cdot p$, and presents a logarithmic mass singularity as $m\to~0$. For $p^\perp=0$ and $b^\pm=0$, the mechanical component satisfies $\langle\mathbb J^{-a}\rangle=-b^a\langle n\cdot\mathbb K\rangle$. For nonzero $b^a$, its classical high-energy behaviour is
\begin{align}
 \langle\mathbb J^{-a}\rangle_{\text{QED}}\big|_{\hbar=0}&\stackrel{\text{H.E.}}{=} -\frac{8\alpha_{\text{cl}}^2b^a(n\cdot p)^2}{3m^4}\int_0^\infty\!\ud\bar\omega\,|E_c(\bar\omega)|^2\,,\\
 \langle\mathbb J^{-a}\rangle_{\text{GR}}\big|_{\hbar=0}&\stackrel{\text{H.E.}}{\sim} -64G^2b^a(n\cdot p)^2\left[\log\!\left(\frac{\Lambda^2(n\cdot p)^2}{m^2}\right)-\frac{11}{6}\right] \int_0^\infty\!\ud\bar\omega\,I_H(\bar\omega)\,.\nonumber
\end{align}
We conclude that the presence of mass singularities in the high energy behaviour of global classical radiative observables is a universal phenomenon. Local classical observables, instead, can be shown to be free of mass singularities at fixed directions with $0<|R^\perp|<\infty$, i.e. away from directions collinear or anticollinear to $\hat n$, under the stated frequency-convergence assumptions, see~(\ref{eq:emitted_P_QED_cl}), (\ref{eq:emitted_N_QED_cl}) for electrodynamics and (\ref{eq:emitted_P_GR_cl}), (\ref{eq:emitted_N_GR_cl}) for gravity\footnote{This mirrors the discussion for the waveform in ~\cite{Kovacs:1977uw,Kovacs:1978eu,DiVecchia:2022nna,Adamo:2022qci}.}. In our wave-particle scattering setup we have been able to see these mass singularities for radiative observables at leading order in perturbation theory, namely at $\alpha^2$ and $G^2$, unlike the analogous wave observables for the classical two-body problem \cite{Saketh:2021sri,Bern:2021xze,Dlapa:2022lmu,Dlapa:2023hsl,DiVecchia:2022nna}.

\subsubsection*{The high-energy limit in electrodynamics to all orders in the coupling}
Staying in the classical theory, and focusing on electrodynamics, we now ask whether higher-order effects (in the coupling) can resolve the mass singularities found above. Here we avail ourselves of the QED literature, which contains \emph{all-orders} results for classical and quantum scattering in plane-wave backgrounds.

Our observables receive two types of corrections, both related to higher powers of the coupling. Higher powers of $eE$ describe (conservative) Lorentz force effects. Each order in $eE$ also receives an infinite series of corrections in $e^2$, which are radiation reaction effects.\footnote{See~\cite{Kosower:2018adc,Elkhidir:2023dco} for some interesting comments on radiation reaction from a pure amplitude perspective.}

The impact of classical radiation and {radiation} reaction on a particle orbit is described by the Lorentz-Abraham-Dirac equation (LAD), which famously suffers from unphysical runaway and pre-acceleration effects, {see~\cite{Coleman:1961zz,Burton:2014wsa} for reviews}. The Landau-Lifshitz (LL) equation~\cite{Landau:1975pou},
\be\label{LL}
	m \ddot{x}_\mu = eF_{\mu\nu}(x)\dot{x}^\nu + \frac23 \frac{e^2}{4\pi} \Big[\frac{e}{m}{\dot F}_{\mu\nu}(x)\dot{x}^\nu  + \frac{e^2}{m^2} F^2_{\mu\nu}(x) \dot{x}^\nu{\dot x}^2 -\frac{e^2}{m^2}\dot{x}_\mu \dot{x}^\nu F^2_{\nu\sigma}(x){\dot x}^\sigma\Big] \;,
\ee
is an approximation to LAD obtained by reduction of order. It is blind to the unphysical features of LAD (as are its generalisations~\cite{Ekman:2021eqc}), but, as argued in e.g.~\cite{Koga,DiPiazza:2018luu,Ekman:2021vwg}, the difference between LAD and LL is terms which are smaller than quantum effects. LL admits an exact solution in plane wave backgrounds~\cite{schrufer,Voloshenko,Piazza2008ExactSO}, and this has recently been re-derived analytically from an all-orders resummation of QED loop diagrams~\cite{Torgrimsson:2021wcj}. Equation~(\ref{LL}) thus allows us to explicitly explore the impact of higher-order and all-order classical effects. (The unphysical problems of radiation reaction as described by LAD are expected to be resolved by quantum effects~\cite{Moniz:1976kr,Johnson:2000qd} -- we will see a manifestation of this idea below.)

As our interest is only in the high-energy and massless limits of these existing results, we will be brief here, referring the reader to~\cite{Piazza2008ExactSO,Ekman:2021vwg} for more complete expressions\footnote{In \cite{Ekman:2021vwg}, arXiv version, eq.(21) gives the solution of the LL equation, in terms of functions defined in eqs.(22)--(23). The radiated energy-momentum $K_\mu$ is computed from the current via the standard classical formula in eq.(28), the final expression in the plane wave case being eq.(30).}. Let $\mathcal{K}^\mu$ denote the radiated momentum computed using the leading local radiation rate, neglecting explicit field-gradient terms in the LL solution. At high energy we find
\be
    \lim_{n\cdot p\to \infty}   \frac{ n\cdot \mathcal{K}}{n\cdot p}  = 1 \;, 
\ee
while the three remaining components are of order $\sim (n\cdot p)^0$. Within this approximation, resummed radiation reaction effects give sensible high-energy behaviour; while $n\cdot \mathcal{K}$ formally diverges as $n\cdot p\to\infty$, it is bounded above by the initial particle energy. However, the mass singularities remain in other components of the radiated momentum. To 
exhibit these compactly, we define quantities $f$ and $g$ which are purely functions of the driving field,
\be
	f(x^\LCm) := \int^{x^\LCm}_{-\infty}\!\ud y^\LCm\, {E_a(y^\LCm)^2} \;, \qquad g_a(x^\LCm) := \int^{x^\LCm}_{-\infty}\!\ud y^\LCm\, f(y^\LCm) E_a(y^\LCm) \;,
\ee
stripped of all factors of the coupling. In terms of these, the high energy part of the radiated momentum behaves as, writing down \emph{only the terms most singular at zero mass}, 
\be\label{K-LL2}
    \mathcal{K}^\mu \stackrel{\text{H.E.}}{\sim} (n\cdot p)\ell^\mu + \frac{e^6}{{m^4}}\frac{1}{12\pi} n^\mu \int\!\ud x\, f'(x)\frac{g_a(x)g_a(x)}{f(x)^2}  - e\delta_\mu^a\int\!\ud x\, f'(x) \frac{g_a(x)}{f(x)^2} \;.
\ee
Observe that, compared to the lowest-order perturbative calculations above, these all-order results diverge with the \emph{same} power of the mass, $1/m^4$, even though the `coefficients' of these divergences are different. These calculations show us that, strictly, the \emph{perturbative limit} and the \emph{massless limit} do not commute. However, the leading power of the mass divergence is the \emph{same} independent of whether we work to leading order in perturbation theory in the coupling, or to all orders in perturbation theory.  This is why, a posteriori, it is sufficient to work to leading perturbative order to exhibit mass divergences, as we did in earlier sections. These mass singularities persist away from the high-energy limit.

The conclusion is that there are no classical effects which can remove the mass divergence of the theory. Thus, if anything can secure a finite massless limit, it must be quantum effects, and we will indeed demonstrate this below. In gravity, while all-orders corrections in $\kappa H$ are available in plane wave backgrounds~\cite{Garriga:1990dp,Adamo:2017nia}, all-orders radiation reaction (or self-force) effects are not. We believe, though, that essentially the same story holds -- there are no classical effects which remove the logarithmic mass divergence seen above.

\subsection{Quantum theory: the resolution for a smooth limit}
We begin by considering \emph{perturbative} quantum corrections to classical results. It is again sufficient to focus on $\langle n\cdot \mathbb{K}\rangle$ to convey the main message. Quantum corrections are obtained simply by retaining higher orders in the $\hbar$-expansion. In electrodynamics we obtain
\be\begin{split}\label{thing1}
	\hspace{-12pt}\langle n\cdot \mathbb{K} \rangle_{\text{QED}} = &8 \alpha_{\text{cl}}^2 \frac{(n \cdot p)^2}{3 m^4}  \int_0^\infty\!\ud {\bar \omega}\, |E_a({\bar \omega})|^2 \bigg[
	1-{\frac{21}{5}}\frac{\hbar {\bar \omega} n\cdot p}{m^2}+\frac{66}{5} \bigg(\frac{\hbar {\bar \omega} n\cdot p}{m^2}\bigg)^2+ \ldots
	\bigg].
\end{split}
\ee
Rather than eliminating the classical mass singularities, each quantum correction also diverges as $n\cdot p\to\infty$ or $m\to 0$! A similar story holds in gravity. However, the conclusion that the quantum theory is sick at $m=0$ is incorrect: there are terms which are \emph{not} captured by a perturbative expansion in powers of $\hbar$, and which yield a finite quantum result.

Assuming that $\int_0^\infty\!\ud\bar\omega\,|E_a(\bar\omega)|^2/\bar\omega^2<\infty$, we return to the full quantum result (\ref{eq:emitted_K_QED}) and take the high-energy limit directly, finding
\be\label{nK-QED-m-0}
\langle n\cdot \mathbb{K}\rangle_{\text{QED}} \stackrel{\text{H.E.}}{\sim} \alpha^2 \int_0^\infty\!\ud \bar{\omega}   \frac{|E_a(\bar{\omega})|^2}{\bar{\omega}^2} \;.
\ee
This is manifestly free of mass singularities. Similarly, in gravity, the leading-order behaviour of the radiated momentum (\ref{eq:emitted_K_GR}) in the high-energy limit is
\be\label{nK-GR-m-is-0}
\langle n\cdot \mathbb{K}\rangle_{\text{GR}} \stackrel{\text{H.E.}}{\sim}
 64 G^2 (n\cdot p)^2\int_0^\infty\!\ud\bar{\omega}\,I_H(\bar\omega) \bigg(\log \bigg[\frac{\Lambda^2n\cdot p}{2 \hbar \bar{\omega}}\bigg]- \frac32\bigg) \;,
\ee
and, at fixed $n\cdot p$ and $\Lambda$, this component has a finite massless limit if
\[
\int_0^\infty\!\ud\bar\omega\,I_H(\bar\omega)\log\!\left(1+\frac{\Lambda^2n\cdot p}{2\hbar\bar\omega}\right)<\infty.
\] The same holds also for the other components of the radiated momentum, as well as for the angular momentum. Indeed, considering the component $J^{-a}$ as before, we have
\begin{align}\label{nJ-m-is-0}
 \langle\mathbb J^{-a}\rangle_{\text{QED}}&\stackrel{\text{H.E.}}{\sim} -\alpha^2b^a\int_0^\infty\!\ud\bar\omega\,\frac{|E_c(\bar\omega)|^2}{\bar\omega^2}\,,\\
 \langle\mathbb J^{-a}\rangle_{\text{GR}}&\stackrel{\text{H.E.}}{\sim} -64G^2b^a(n\cdot p)^2\int_0^\infty\!\ud\bar\omega\,I_H(\bar\omega) \left[\log\!\left(\frac{\Lambda^2(n\cdot p)}{2\hbar\bar\omega}\right)-\frac32\right]\,,\nonumber
\end{align}
which have a smooth massless limit.  Thus we find that the quantum theory has regulated completely the mass singularities of the classical theory. 

\begin{figure}[t!]
    \centering
    \includegraphics[width=0.9\textwidth]{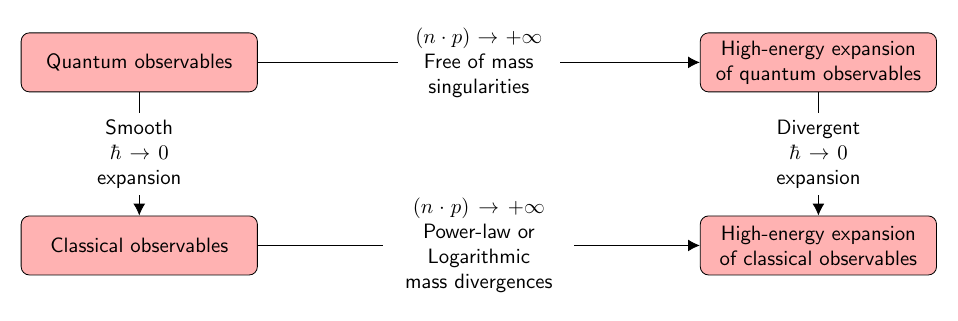} 
    \caption{A schematic representation of the behaviour of classical and quantum global observables in electrodynamics and gravity, which shows their high-energy behaviour and highlights the non-commutativity of different limits. }
    \label{fig:diagram_scheme}
\end{figure}

In summary, while one can take the high energy or massless limit in the full quantum theory, as in (\ref{nK-QED-m-0}) and (\ref{nK-GR-m-is-0}), quantum effects are large in this regime, as signalled by inverse powers or logs of~$\hbar$. These forbid the classical limit from being taken\footnote{Radiation emitted by massless charged particles has previously been identified as fully quantum in~\cite{Galtsov:2015rcs}.}, which explains why the classical mass divergences were not resolved by adding \emph{perturbative} quantum corrections. Starting in the classical theory, on the other hand, the high-energy limit means harder collisions and more radiation, and classical observables become large (formally diverge), but in the same limit the classical wavelength $\lambdabar$ becomes small relative to the Compton wavelength, $ \lambdabar/\lambdabar_C \ll 1$. This means that quantum effects are, naturally, \emph{large}. 
Thus trying to take, in either order, the limits $m \to 0$ and $\hbar\to 0$ leads to a divergence. This is in-line with literature results for the \emph{conservative} observables of~\cite{Bern:2019crd} where the origin of mass singularities was traced back to the inability to interchange the classical and massless limits. The situation is summarised in Fig.~\ref{fig:diagram_scheme}.

A comment is in order on the relation of our findings to analogous discussions of the classical two-body problem in \cite{Bern:2019crd,DiVecchia:2022nna,DEath:1976bbo,Kovacs:1977uw,Kovacs:1978eu,Gruzinov:2014moa}. In our case the impact parameter $b$ plays a different role because of the symmetries of the plane-wave; moreover we have performed our calculations perturbatively (see the inequalities~\eqref{eq:CED_classical_observables} and \eqref{eq:GR_classical_observables}). Nevertheless, a comparison between the functions arising in the high-energy limit of~\cite{Saketh:2021sri,Bern:2021xze} and \cite{Herrmann:2021tct,Mougiakakos:2021ckm,Jakobsen:2021smu,DiVecchia:2021bdo,DiVecchia:2022nna,Dlapa:2022lmu,Dlapa:2023hsl,Bini:2022enm,Damour:2022ybd} and their dependence on the rapidity $\gamma' = (p_1 \cdot p_2)/ (m_1 m_2)$ shows  remarkable similarities with our results expressed in terms of our $\gamma = (n \cdot p) /m$. It is tempting to suggest that the quantization of the electromagnetic (and gravitational) field is eventually needed if we are interested in obtaining radiative observables\footnote{A complementary analysis for conservative observables has instead been performed in \cite{Bern:2019crd}.} free of mass singularities in perturbation theory, as shown earlier with our analysis of classical and quantum results.

\section{Conclusion and outlook}

The space and analytic structure of wave scattering observables is not only of theoretical interest, but also relevant for physical applications from collider to gravitational wave physics. In this paper, we have explored the space and properties of local and global observables for the radiation emitted by a scalar moving in a plane-wave background both in quantum electrodynamics and in general relativity treated as an effective field theory.  Such plane-wave profiles have the interesting feature that some observables can be computed exactly at all orders in perturbation theory, which makes them an ideal playground for addressing some of the important questions which arise in the study of the two-body problem.

In this paper we have been interested in understanding both global and local angle-dependent properties of the emitted radiation. Just as the differential cross-section is the local analogue of the cross-section, we have studied the local versions of the radiated momentum and angular momentum. We began by discussing gauge invariant representations for the operators corresponding to the momentum and angular momentum flow, whose expectation value is closely related to the waveform profile structure at infinity. Indeed, we showed that those local observables are completely determined by the amplitude and the phase of the radiative waveform, at least for a coherent state profile.

We then proceeded to compute, in perturbation theory, the radiated momentum and angular momentum, as well as their local analogues, first in the quantum theory and then in the classical limit. The leading contribution is given by integrating a tree-level Compton-like amplitude and its conjugate against an operator kernel related to the specific observable. We saw that, in gravity, the collinear divergence of the gravitational Compton cross-section poses conceptual challenges for the calculations of global observables. We have analysed this fundamental issue and shown with the KLN theorem that summing over degenerate forward-scattered gravitons is required to formally achieve an infrared-finite cross-section, thus generalising results of~\cite{Frye:2018xjj} from gauge theory to gravity. This motivated our proposal to dress the incoming state in the collinear direction, in order to define gravitational observables integrated over the celestial sphere. With this prescription, we found new compact expressions for our observables at order $\alpha^2$ in quantum electrodynamics and $G^2$ in the gravitational effective theory. In our overlap with the literature, we find complete agreement. 

We observed that, on general grounds, the high-energy limit of our classical radiative observables generically presents power-law mass singularities in QED and logarithmic mass singularities in GR. This resonates with recent results obtained for the classical observables of the two-body problem, and allows us to offer an explanation of the puzzle and its resolution in the simpler context of wave-particle scattering. These mass singularities eventually arise when we scatter objects with wavelengths larger than the Compton wavelength in a classical context, and can only be fully resolved within the quantum theory, as we showed explicitly. In particular, there are large quantum corrections to observables in the high-energy region, which manifest themselves in $1/\hbar$ effects which cannot have a classical interpretation. Moreover, thanks to the all-order results available in the `strong field QED' literature for the energy emitted in plane-wave backgrounds, we can see in such cases that resumming all-order radiation effects cannot cure the mass singularity, and a smooth massless limit can only be achieved within the quantum theory. Interestingly, a related problem in electrodynamics on the overestimation of energy emitted in the classical scattering process is also solved by quantum effects; see~\cite{Blackburn:2019rfv,Gonoskov:2021hwf} for reviews, and \cite{Cole:2017zca,Poder:2017dpw} for experimental results.

There are many open avenues for future research. It is not yet clear how many collider QCD observables, mainly developed for jet physics, can be imported into the gravitational context. It would be interesting to explore this further for the full two-body problem, as it might offer a new perspective for gravitational wave physics observables and their all-order resummation. Moreover, it would be nice to provide a complementary derivation of the observables discussed here from a CFT point of view \cite{Hofman:2008ar,Kologlu:2019mfz,Lee:2022ige,Caron-Huot:2022eqs,Hu:2022txx}, where questions like IR-finiteness can be answered in a more straightforward way through their non-perturbative definition \cite{Belitsky:2001ij,Belitsky:2013bja,Belitsky:2013xxa,Chen:2020vvp,Korchemsky:2021okt}.

\paragraph{Acknowledgements.} We thank T.~Adamo, N.E.J.~Bjerrum-Bohr, A.~Cristofoli, G.~Korchemsky, D.~Kosmopoulos, S.~Klisch, M.~Lavelle, D.~McMullan, R.~Stegeman, G.~Torgrimsson, M.~Zeng and S.~Zhiboedov for insightful discussions. We are extremely grateful to H.~Hannesdottir and M.~Schwartz for useful discussions on the KLN theorem \cite{Frye:2018xjj} and for comments on the draft, as well as to C.~Heissenberg for interesting comments on the high-energy limit and for a critical reading of this manuscript. A.I. thanks Polux Gabriel Garcia Elizondo for useful discussions on the literature. This research was supported by the National Science Foundation under Grant No. NSF PHY-1748958 (RG) and the STFC consolidator grant ST/X000494/1 ``Particle Theory at the Higgs Centre'' (AI).

\appendix

\section{Perturbation theory in the EFT approach to quantum gravity}
\label{sec:appendixA}

Here we collect the perturbative Feynman rules \cite{Rafie-Zinedine:2018izq} which we use for the one-loop calculation in the EFT approach to quantum gravity in section \ref{sec:KLNtheorem}. The propagators of our fields in momentum space in $d$ dimensions are
\begin{align}
G_{\phi}(p) &= \frac{i}{p^2 - m^2 + i \epsilon} \,, \qquad \qquad \qquad G_{\chi}^{\mu \nu}(p) = -\frac{i \eta^{\mu \nu}}{p^2 + i \epsilon} \,,  \nonumber \\
G_h^{\mu \nu \alpha \beta}(p) &= \frac{i P^{\mu \nu \alpha \beta}}{p^2 + i \epsilon} \,, \qquad P^{\mu \nu \alpha \beta} = \frac{1}{2} \left(\eta^{\mu \alpha} \eta^{\nu \beta} +  \eta^{\mu \beta} \eta^{\nu \alpha} - \frac{2}{d-2} \eta^{\mu \nu} \eta^{\alpha \beta} \right) \,.
\end{align}
The interaction lagrangian we use is:
\begin{align}
\mathcal{L}_{h h h} &=\frac{\kappa}{2}\left(\frac{1}{4} h_\mu{ }^\mu \partial_\nu h_\alpha{ }^\alpha \partial^\nu h_\beta{ }^\beta-h^{\mu \nu} \partial_\mu h^{\alpha \beta} \partial_\nu h_{\alpha \beta} +2 h^{\mu \nu} \partial_\mu h^{\alpha \beta} \partial_\alpha h_{\nu \beta}-\frac{1}{2} h^{\mu \nu} \partial_\alpha h_{\mu \nu} \partial^\alpha h_\beta{ }^\beta\right) \,, \nonumber \\
\mathcal{L}_{h h h h} &=\frac{\kappa^2}{4}\Big( -\frac{5}{16} h_\mu{ }^\mu h_\nu{ }^\nu \partial_\alpha h_\beta{ }^\beta \partial^\alpha h_\tau{ }^\tau+\frac{1}{2} h_\mu{ }^\mu h^{\nu \alpha} \partial_\nu h_{\beta \tau} \partial_\alpha h^{\beta \tau}  -h_\mu{ }^\mu h^{\nu \alpha} \partial_\nu h^{\beta \tau} \partial_\beta h_{\alpha \tau}  \nonumber\\
& \qquad \qquad +h_\mu{ }^\mu h^{\nu \alpha} \partial_\beta h_{\nu \tau} \partial^\beta h_\alpha{ }^\tau-\frac{1}{8} h_{\mu \nu} h^{\mu \nu} \partial_\alpha h_\beta{ }^\beta \partial^\alpha h_\tau{ }^\tau+h^{\mu \nu} \partial_\mu h_{\nu \alpha} \partial^\beta h^{\alpha \tau} h_{\beta \tau}  \nonumber\\
& \qquad \qquad +\frac{1}{4} h^{\mu \nu} \partial_\mu h_\alpha{ }^\alpha h_{\nu \beta} \partial^\beta h_\tau{ }^\tau-2 h^{\mu \nu} \partial_\mu h^{\alpha \beta} h_{\nu \alpha} \partial^\tau h_{\beta \tau} +h^{\mu \nu} \partial_\mu h_{\alpha \beta} h_{\nu \tau} \partial^\tau h^{\alpha \beta} \nonumber \\
& \qquad \qquad -2 h^{\mu \nu} \partial_\mu h_{\alpha \beta} h^{\alpha \tau} \partial_\tau h_\nu{ }^\beta+h^{\mu \nu} h_{\nu \alpha} \partial_\beta h_{\mu \tau} \partial^\beta h^{\alpha \tau}+2 h^{\mu \nu} \partial_\nu h_{\alpha \beta} h^{\alpha \beta} \partial^\tau h_{\mu \tau}\Big)\,, \nonumber \\
\mathcal{L}_{\phi \phi h} &=\frac{\kappa}{2}\left(-\frac{1}{2} h_\mu^\mu \phi^2 m^2+\frac{1}{2} h_\mu^\mu \partial_\nu \phi \partial^\nu \phi-h^{\mu \nu} \partial_\mu \phi \partial_\nu \phi\right)\,, \nonumber \\
\mathcal{L}_{\phi \phi h h} &=\frac{\kappa^2}{4}\left(h^{\mu \nu} h_\nu{ }^\alpha \partial_\mu \phi \partial_\alpha \phi-\frac{1}{2} h_\mu{ }^\mu h^{\nu \alpha} \partial_\nu \phi \partial_\alpha \phi\right) \,, \nonumber \\
\mathcal{L}_{\phi \phi h h h}  & =\frac{\kappa^3}{8} \Big( -\frac{1}{4} m^2 \phi^2 h_\mu{ }^\mu h^{\nu \alpha} h_{\nu \alpha}-\frac{1}{16} \phi \partial_\mu \phi \partial^\mu h_\nu{ }^\nu h_\alpha{ }^\alpha h_\beta{ }^\beta  -\frac{1}{2} \phi \partial_\mu \phi \partial^\mu h^{\nu \alpha} h_\nu{ }^\beta h_{\alpha \beta} \nonumber \\
&\qquad \qquad +\frac{1}{4} \partial_\mu \phi \partial^\mu \phi h_\nu{ }^\nu h^{\alpha \beta} h_{\alpha \beta} +\frac{1}{8} \partial_\mu \phi \partial_\nu \phi h^{\mu \nu} h_\alpha{ }^\alpha h_\beta{ }^\beta-\frac{1}{2} \partial_\mu \phi \partial^\nu \phi h^{\mu \alpha} h_{\nu \alpha} h_\beta{ }^\beta \nonumber \\
&\qquad \qquad-\partial^\mu \phi \partial^\nu \phi h_{\mu \alpha} h_{\nu \beta} h^{\alpha \beta}\Big) \,, \nonumber \\
\mathcal{L}_{\bar{\chi} \chi h}&=\frac{\kappa}{2} \Big( -\bar{\chi}^\mu \chi^\nu \partial_\mu \partial_\nu h_\alpha{ }^\alpha+\bar{\chi}_\mu \partial^\mu \chi_\nu \partial_\alpha h^{\nu \alpha}+2 \bar{\chi}_\mu \chi^\nu \partial_\nu \partial_\alpha h^{\mu \alpha} -\frac{1}{2} \bar{\chi}_\mu \partial_\nu \chi^\nu \partial^\mu h_\alpha{ }^\alpha \nonumber \\
&\qquad \qquad -\bar{\chi}_\mu \partial_\nu \chi_\alpha \partial^\mu h^{\nu \alpha}+\bar{\chi}_\mu \partial_\nu \chi_\alpha \partial^\alpha h^{\mu \nu}  -\partial^\mu \bar{\chi}_\mu \partial_\nu \chi_\alpha h^{\nu \alpha}-\partial^\mu \bar{\chi}^\nu \partial_\mu \chi_\nu h_\alpha{ }^\alpha \nonumber \\
&\qquad \qquad -\partial_\mu \bar{\chi}_\nu \partial^\mu \chi_\alpha h^{\nu \alpha} +\partial_\mu \bar{\chi}_\nu \partial_\alpha \chi^\nu h^{\mu \alpha}-\partial_\mu \bar{\chi}_\nu \partial_\alpha \chi^\alpha h^{\mu \nu}\Big)\,, \nonumber \\
\mathcal{L}_{\bar{\chi} \chi h h} &=\frac{\kappa^2}{8} \Big(-\bar{\chi}^\mu \partial^\nu \chi_\nu h_{\mu \alpha} \partial_\beta h^{\alpha \beta}+\bar{\chi}_\mu \partial^\mu \chi^\nu h_{\nu \alpha} \partial_\beta h^{\alpha \beta}+2 \bar{\chi}_\mu \partial^\mu \chi^\nu \partial_\nu h^{\alpha \beta} h_{\alpha \beta} \nonumber \\
&\qquad \qquad +\bar{\chi}_\mu \partial^\mu \chi^\nu \partial^\alpha h_{\nu \alpha} h_\beta{ }^\beta+2 \bar{\chi}_\mu \chi^\nu h^{\mu \alpha} \partial_\nu \partial_\alpha h_\beta{ }^\beta+\bar{\chi}^\mu \chi^\nu \partial_\mu h_\alpha{ }^\alpha \partial_\nu h_\beta{ }^\beta \nonumber \\
&\qquad \qquad +2 \bar{\chi}^\mu \chi^\nu \partial_\nu h_{\mu \alpha} \partial^\alpha h_\beta{ }^\beta+2 \bar{\chi}^\mu \chi^\nu \partial_\nu \partial^\alpha h_{\mu \alpha} h_\beta{ }^\beta+\bar{\chi}^\mu \partial^\nu \chi_\mu h_{\nu \alpha} \partial_\beta h^{\alpha \beta}  \nonumber \\
&\qquad \qquad +\bar{\chi}_\mu \partial^\mu \chi_\nu h^{\nu \alpha} \partial_\alpha h_\beta{ }^\beta-\bar{\chi}^\mu \partial^\nu \chi_\nu \partial_\alpha h_{\mu \beta} h^{\alpha \beta}+\bar{\chi}^\mu \partial^\nu \chi^\alpha h_{\mu \nu} \partial_\alpha h_\beta{ }^\beta \nonumber \\
&\qquad \qquad +\bar{\chi}_\mu \partial_\nu \chi^\alpha h^{\mu \nu} \partial^\beta h_{\alpha \beta}-\bar{\chi}^\mu \partial^\nu \chi^\alpha \partial_\mu h_{\nu \alpha} h_\beta{ }^\beta-\bar{\chi}^\mu \partial^\nu \chi^\alpha h_{\mu \alpha} \partial_\nu h_\beta{ }^\beta \nonumber \\
&\qquad \qquad +\bar{\chi}^\mu \partial^\nu \chi^\alpha h_{\mu \alpha} \partial^\beta h_{\nu \beta}-\bar{\chi}^\mu \partial^\nu \chi^\alpha \partial_\mu h_\beta{ }^\beta h_{\nu \alpha}-\bar{\chi}^\mu \partial^\nu \chi^\alpha \partial_\nu h_{\mu \alpha} h_\beta{ }^\beta \nonumber \\
&\qquad \qquad -4 \bar{\chi}^\mu \partial^\nu \chi^\alpha h_{\nu \alpha} \partial^\beta h_{\mu \beta}+\bar{\chi}^\mu \partial^\nu \chi^\alpha \partial_\alpha h_{\mu \nu} h_\beta{ }^\beta-\partial^\mu \bar{\chi}_\mu \partial^\nu \chi^\alpha h_{\nu \alpha} h_\beta{ }^\beta \nonumber \\
&\qquad \qquad -\partial^\mu \bar{\chi}^\nu \partial_\mu \chi_\nu h^{\alpha \beta} h_{\alpha \beta}-2 \partial^\mu \bar{\chi}^\nu \partial_\mu \chi^\alpha h_{\nu \alpha} h_\beta{ }^\beta+\partial^\mu \bar{\chi}^\nu \partial^\alpha \chi_\nu h_{\mu \alpha} h_\beta{ }^{\beta} \nonumber \\
&\qquad \qquad -\partial^\mu \bar{\chi}^\nu \partial^\alpha \chi_\alpha h_{\mu \nu} h_\beta{ }^\beta-\partial^\mu \bar{\chi}_\nu \partial^\alpha \chi_\alpha h_{\mu \beta} h^{\nu \beta}-4 \partial^\mu \bar{\chi}^\nu \partial^\alpha \chi^\beta h_{\mu \nu} h_{\alpha \beta} \nonumber \\
&\qquad \qquad +2 \partial^\mu \bar{\chi}^\nu \partial^\alpha \chi^\beta h_{\mu \alpha} h_{\nu \beta}+2 \partial^\mu \bar{\chi}^\nu \partial^\alpha \chi^\beta h_{\mu \beta} h_{\nu \alpha} \Big) \,,
\end{align}
and we refer to \cite{Rafie-Zinedine:2018izq} for explicit details about the field redefinition and the gauge choice.

\bibliographystyle{JHEP}
\bibliography{references}
\end{document}